\documentclass[%
 reprint,
superscriptaddress,
 amsmath,amssymb,
 aps,
]{revtex4-2}

\usepackage{graphicx}
\usepackage{dcolumn}
\usepackage{bm}
\usepackage{hyperref}
\usepackage{booktabs}

\begin{document}

\preprint{APS/123-QED}

\title{Merging high localization and TE-TM polarization degeneracy of \\ guided waves in dielectric metasurfaces}

\author{Rui Li}
\affiliation{State Key Laboratory of Integrated Optoelectronics, College of Electronic Science and Engineering, International Center of Future Science, Jilin University, 2699 Qianjin Street, 130012 Changchun, China
}%
\author{Sergey Polevoy}%
\affiliation{Department of Radiospectroscopy, O. Ya. Usikov Institute for Radiophysics and Electronics of the NAS of Ukraine, 61085 Kharkiv, Ukraine
}%
\author{Vladimir Tuz}%
\affiliation{State Key Laboratory of Integrated Optoelectronics, College of Electronic Science and Engineering, International Center of Future Science, Jilin University, 2699 Qianjin Street, 130012 Changchun, China
}%
\affiliation{School of Radiophysics, Biomedical Electronics and Computer Systems, V.~N.~Karazin Kharkiv National University, 4 Svobody Square, 61022 Kharkiv, Ukraine
}%
\author{Oleh Yermakov}%
\email{oe.yermakov@gmail.com}
\affiliation{Department of Computational Physics, V.~N.~Karazin Kharkiv National University, 4 Svobody Square, 61022 Kharkiv, Ukraine
}%
\affiliation{Department of Fiber Photonics, Leibniz Institute of Photonic Technology, 9 Albert-Einstein-Stra{\ss}e, 07745 Jena, Germany
}%

\date{\today}

\begin{abstract}
The polarization degree of freedom is an inherent feature of plane waves propagating in an isotropic homogeneous medium. The miniaturization of optical systems leads to the high localization of electromagnetic waves, but also to the loss of polarization control, namely, breaking TE-TM polarization degeneracy. In this work, we discover the near-field polarization degree of freedom for highly localized guided waves propagating along a dielectric metasurface. We demonstrate the opportunity to create a metasurface with the degenerate TE-TM polarization spectrum for the required operating wavelength and different constitutive materials. In particular, we analyze several possible implementations including silicon nitride and ceramic metasurfaces consisting of disk-shaped resonators, and evaluate the impact of substrate. Finally, we experimentally implement one of the metasurface designs and verify its broadband degenerate TE-TM polarization spectrum. The obtained results form a fundamentally new platform for the planar polarization devices utilizing the polarization degree of freedom of localized light. 
\end{abstract}

\maketitle


\section{Introduction}

\begin{figure}[t]
\centering
\includegraphics[width=1.0\linewidth]{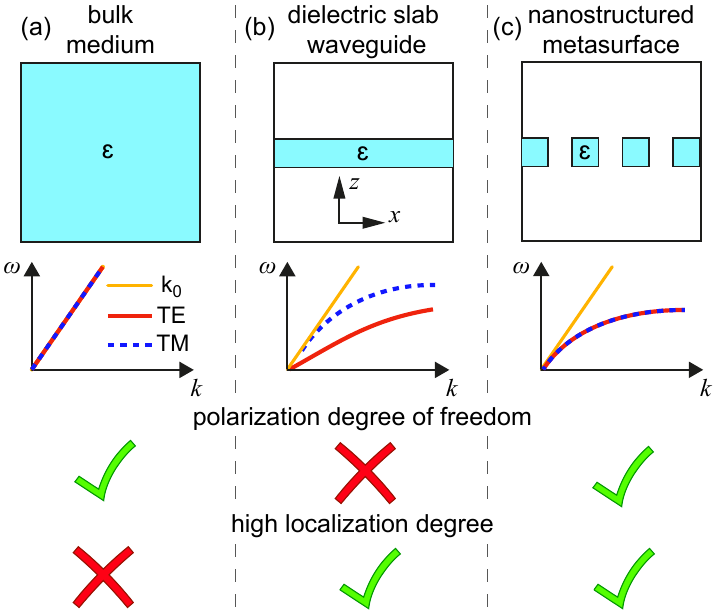}
\caption{Schematic view of an optical system (top line), related dispersion characteristics (second line), presence of the TE-TM ($k_s = k_p$) degeneracy (third line) and high localization ($k > n_0 k_0$) (bottom line) for (a) plane waves in a bulk medium (left column), (b) guided waves in a dielectric slab waveguide (central column), and (c) guided waves in a nanostructured dielectric metasurface (right column), respectively.}
\label{fig:fig_1}
\end{figure}

Modes of a bulk isotropic medium are double-degenerate states with respect to polarization. This means that the dispersion characteristics of the orthogonally polarized transverse electric (TE) and transverse magnetic (TM) modes are the same in such a medium, i.e. their group velocities and wavenumbers completely coincide at any frequency and in any direction, resulting in the mode polarization TE-TM degeneracy [Fig.~\ref{fig:fig_1}(a)]. Therefore, the condition $k_p = k_s = n_0 k_0$ is met for the wave numbers of degenerate modes, where $k_0 = 2 \pi/\lambda$ is the wave number of a plane wave in free space, $\lambda$ is the wavelength, $n_0$ is the refractive index of the medium, and the subscripts $s$ and $p$ distinguish the TE and TM polarizations of the modes, respectively. In general, one can reach any (linear, circular, or elliptical) polarization of waves propagated in an isotropic medium by controlling the corresponding complex amplitudes in the superposition of the TE and TM modes. 

The TE-TM degeneracy allows one to preserve the polarization of an electromagnetic wave when it is propagated through an optical guiding system. Nevertheless, this degeneracy is lifted off in an anisotropic medium constituting the operational principle of classical polarization converters. In such devices, plates of an anisotropic (typically, uniaxial) crystal are widely used. By controlling the difference between the propagation constants $\Delta k = | k_{p} - k_{s} |$ and optical path length $L$ within the anisotropic medium, one can adjust any phase delay $\varphi = \Delta k L$ between the TE and TM modes. Two well-known examples are the quarter-wave and half-wave plates that bring the phase delay of $\varphi = \pi/2$ and $\varphi = \pi$, respectively. For an incident wave with linear diagonal polarization, the output of the quarter-wave plate is a wave with circular polarization, and a wave with linear diagonal polarization rotated by 90$^\circ$ is the output of the half-wave plate \cite{born2013principles}. To get a tunable optical device, one needs to manipulate the polarization of electromagnetic waves at the system output,  i.e. be able to switch between degenerate and non-degenerate states of polarization. This switching mechanism can be implemented for plane waves in a medium by changing its properties by some external influence from isotropic to anisotropic \cite{Mu_PhotonRes_2019, Xiao_PhysRevApplied_2024, Zhu_PhotonRes_2024}. As anisotropic media in which the TE and TM modes are non-degenerate, uniaxial and biaxial crystals \cite{collett2005field}, waveguides \cite{ChavezBoggio_JOSAB_2014, Zhao_PhotonRes_2024}, multilayer structures \cite{Fesenko_Nanophot_2016, Fesenko_OptCommun_2016}, metamaterials \cite{Benedikovic_OptExpress_2017}, and metasurfaces \cite{Teng_PhotonRes_2019, Pu_AdvOptMater_2019, Xiao_PhysRevResearch_2024} are used in optical systems (see also comprehensive reviews \cite{Garanovich_PhysRep_2012, Halir_lpor_2014, Quaranta_lpor_2018, Meng_Light_2021} and references therein).

The modern trends in miniaturization and planarization of on-chip devices, optoelectronic and integrated circuits require control over highly localized electromagnetic waves, such as guided and surface waves. The electromagnetic field of the localized modes exponentially decays in the direction orthogonal to the propagation direction. In this case, the dispersion characteristics of the TE and TM modes lie below the light line in a bulk isotropic medium, in the so-called near-field or dark-field region, where the wave numbers of the modes $k_{s,p}$ are larger than those of plane waves in a bulk isotropic medium, $k_{s,p} > n_0 k_0$. The transition from the plane waves in bulk media to the localized guided waves in waveguides provides strong field confinement but also lifts off the TE-TM degeneracy of modes. Namely, for the localized waves, the dispersion characteristics of the TE and TM modes do not coincide in a general case. The simplest example of such a system supporting non-degenerate localized modes is a dielectric slab waveguide [see Fig.~\ref{fig:fig_1}(b)]. Nevertheless, in such structures, it is possible to achieve polarization degeneracy by using the waveguides with symmetric cross-sections~\cite{snyder1983optical} and by utilizing anisotropic nanostructures with the accidental polarization degeneracy related to the intersections between the different dispersion curves~\cite{yermakov2016spin}. The first exception allows the transferring of electromagnetic waves only along a single direction, the second one brings the TE-TM degeneracy for modes only in several narrow spectral bands. In fact, the absence of broadband and multi-directional TE-TM degeneracy severely limits the potential applications of planar optical and photonic devices.

\begin{figure}[t]
\centering
\includegraphics[width=1.0\linewidth]{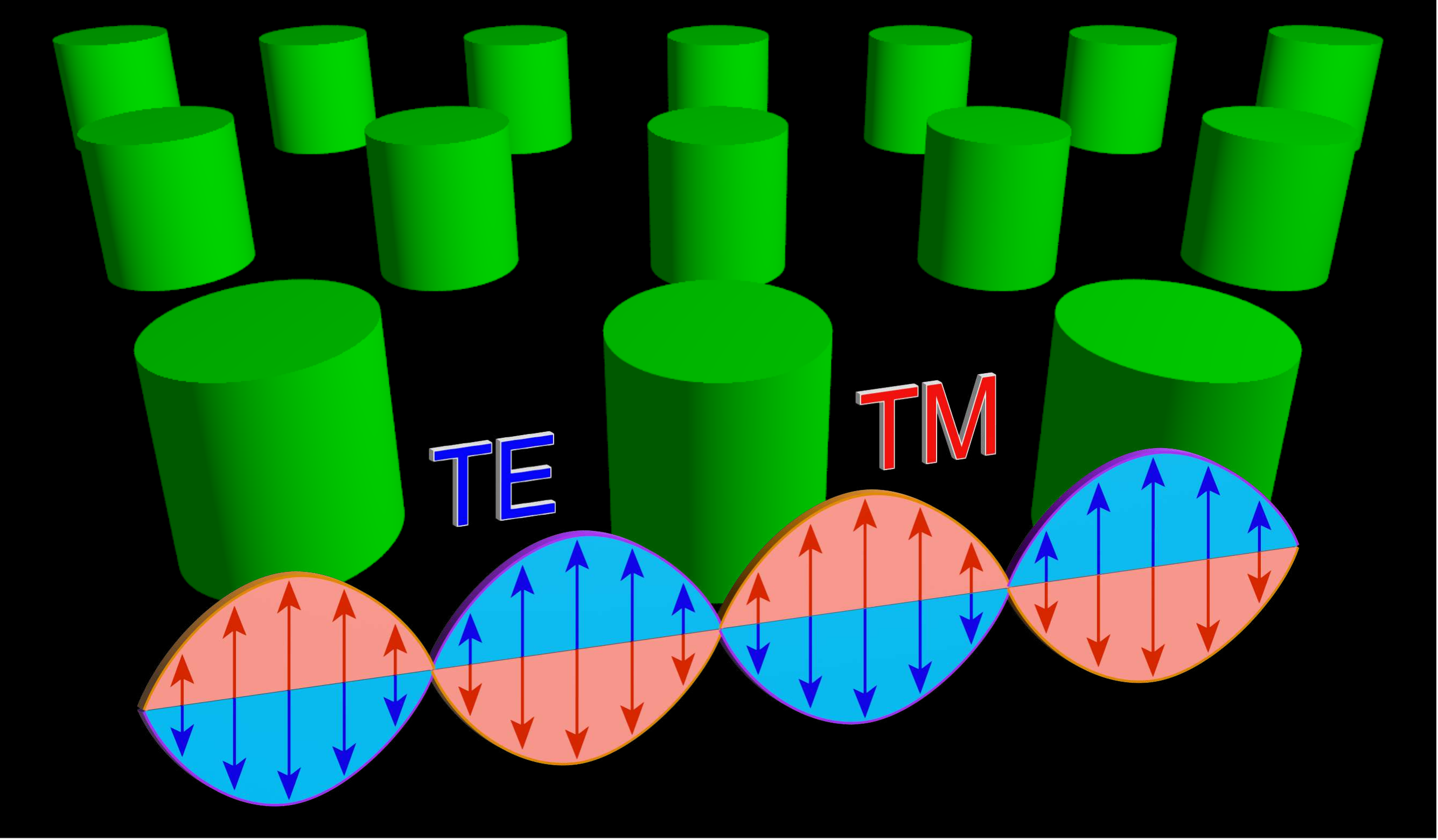}
\caption{Schematic view of the guided TE and TM modes propagating along the dielectric metasurface with the same group and phase velocities.}
\label{fig:fig_art}
\end{figure}

This paper aims to demonstrate the principle of designing an optical platform capable of supporting localized guided waves with TE-TM degeneracy (Fig.~\ref{fig:fig_art}). In the simplest approximation, it is implemented by structuring a dielectric slab waveguide [see Fig.~\ref{fig:fig_1}(c)]. More specifically, we engineer the collective Mie-type modes~\cite{evlyukhin2012demonstration, kuznetsov2012magnetic, kruk2017functional} of a dielectric metasurface which is a periodic array of disk-shaped high-refractive-index particles (resonators). In particular, we present a design of a system bearing the TE- and TM-guided modes, whose dispersion characteristics are as close as possible to each other. Unlike some previous studies \cite{Yermakov_PhysRevX.11.031038,Asadulina_lpor_2023}, where a single-directional TE-TM degeneracy was achieved, here we make a generalization considering two-dimensional structures (metasurfaces) made of different materials and operated in different wavelength ranges exhibiting the multidirectional TE-TM degeneracy. To show the generality of our principle, we consider several particular designs of the metasurface made of silicon nitride disks for their operation in the visible and near-infrared (NIR) range. Then, we analyze the impact of the presence of a substrate and confirm that the TE-TM degeneracy can be achieved for structures deposited on a substrate made of material with a low refractive index, e.g. quartz. Finally, we verify our findings with the quasi-optical (microwave) experiment by demonstrating the propagation of guided modes with high polarization degeneracy for all in-plane directions in a metasurface made of low-loss ceramic resonators. The obtained results discover the near-field polarization degree of freedom and introduce a novel platform for planar photonic polarization devices, opening new horizons for many applications including polarization converters, filters, demultiplexers, optical isolators, and sensors.

\section{Theoretical Description}
\subsection{Problem Statement}

In what follows, we consider a metasurface consisting of identical disk-shaped dielectric resonators with a diameter $d$, height $h$, and refractive index $n$. The disks are arranged in a square array with a lattice constant $a$. The superstrate is air ($n_\textrm{sup} = 1$), whereas the substrate is made of an isotropic non-magnetic dielectric material characterized by the refractive index $n_\textrm{sub}$. The thickness of the substrate is assumed to be much greater than the height of the disks, i.e. it is considered to be semi-infinite.

There is a plethora of materials and techniques for fabricating waveguide structures depending on their required operating wavelength \cite{Baranov_Optica_2017, Su_admt_2020}. In particular, as a material for resonators, one can consider zinc oxide (ZnO) \cite{aguilar2019optoelectronic}, silicon nitride (Si$_3$N$_4$) \cite{beliaev2022optical}, titanium dioxide (TiO$_2$) \cite{sarkar2019hybridized} for the visible and near-infrared ranges and silicon (Si) \cite{schinke2015uncertainty, shkondin2017large} for the infrared range. The following requirements determine the choice of the best material: (i) negligible absorption losses, (ii) high refractive index, and (iii) weak material dispersion in the corresponding wavelength range. 

\begin{figure*}[t]
\centering
\includegraphics[width=0.85\linewidth]{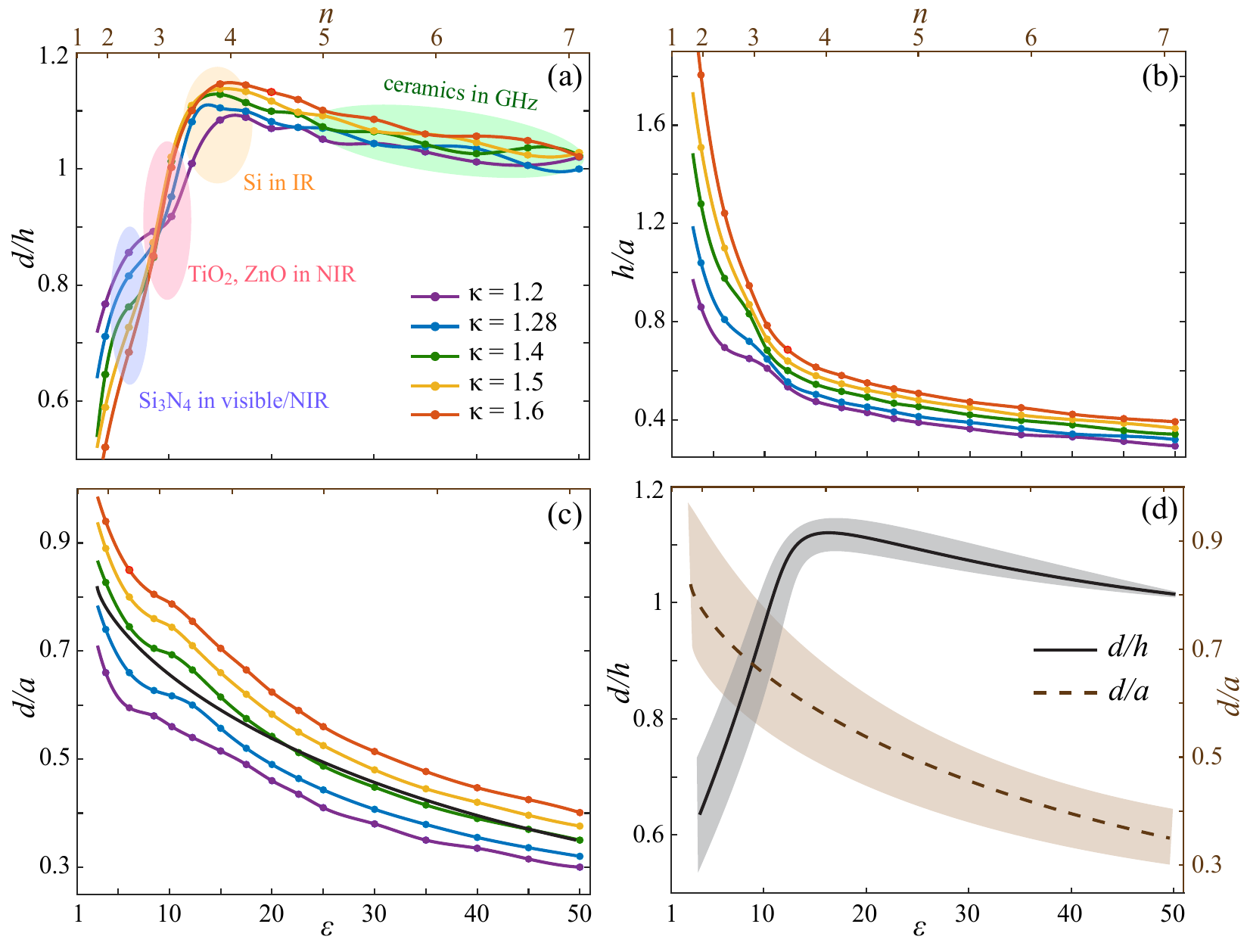}
\caption{Parametric dependencies of (a)  diameter-to-height $d/h$, (b) height-to-period $h/a$ and (c) diameter-to-period $d/a$ ratios on the permittivity (bottom scale) and refractive index (upper scale) of the constitutive material of the disks for different LDs [purple: $\kappa = 1.2$ ($Z_\lambda = 0.12$), blue: $\kappa = 1.28$ ($Z_\lambda = 0.1$), green: $\kappa = 1.4$ ($Z_\lambda = 0.081$), yellow: $\kappa = 1.5$ ($Z_\lambda = 0.071$), red: $\kappa = 1.6$ ($Z_\lambda = 0.064$)]. The dots and lines correspond to the numerically calculated optimized designs and corresponding fitting distributions. The colored regions in panel (a) correspond to different materials and spectral ranges. (d) Material dependencies of the universal $d/h$ (left scale) and $d/a$ (right scale) ratios corresponding to the related values averaged over different LDs.}
\label{fig:fig_2}
\end{figure*}

The important value characterizing the TE-TM degeneracy is the difference between the wave numbers (DBW) of the TM ($k_p$) and TE ($k_s$) modes at the same wavelength:
\begin{equation}
    \Delta k = |k_p - k_s|.
\end{equation}
This value directly shows how fast the phase delay increases during the wave propagation in a given medium at a given wavelength. We also introduce the distance $L_\lambda$ at which the guided wave propagating along the metasurface with the specified DBW acquires the phase delay $\varphi_0 = 0.1\pi$ normalized per the wavelength $\lambda$:
\begin{equation}
    L_\lambda = \frac{L}{\lambda} = \frac{\varphi_0}{\Delta k \lambda} = \frac{a}{10 \lambda} \left( \Delta k a / \pi \right)^{-1}.
\end{equation}
This value characterizes the distance at which the propagating guided wave preserves the polarization state close to the initial condition.

Another critical value characterizing the localization degree (LD) of a guided wave is defined as the ratio between the guided mode and plane-wave wave numbers:
\begin{equation}
    \kappa = \frac{k}{n_\textrm{sub} k_0},
\end{equation}
At the boundary of the first Brillouin zone ($k = \pi/a$), the localization reaches its maximum $\kappa_{max} = \lambda_B/(2 n_\textrm{sub} a)$, where $\lambda_B$ is the guided-mode wavelength at the resonance. In contrast, the phase velocity of the guided wave reaches its minimum. We also introduce the LD as the decay length $Z$ of the surface wave in the direction perpendicular to the metasurface plane normalized per the operating wavelength $\lambda$:
\begin{equation}
    Z_{\lambda} = \frac{Z}{\lambda}  = \frac{1}{2 k_z\lambda} = \frac{1}{4 \pi n_\textrm{sub} \sqrt{\kappa^2 - 1}},
\end{equation}
where $k_z = \sqrt{k^2 - n_\textrm{sub}^2 k_0^2}$. The normalized decay length $Z_\lambda$ means the distance from the metasurface plane at which the field intensity decreases in $e$ times in the units of the incident wavelength.

The maximal LD and the average/maximal DBW within a wide wavelength range are directly proportional. This means it is more difficult to achieve the TE-TM degeneracy (low values of DBW) for a higher level of localization (high values of LD) and vice versa. As mentioned above (Fig.~\ref{fig:fig_1}), while the bulk medium has perfect TE-TM mode degeneracy for non-localized waves, and the guided waves in the dielectric plate are well localized but have no polarization degeneracy, the optimal result in the TE-TM degeneracy control can be obtained in dielectric metasurfaces as an intermediate platform. Such structures can support guided waves with higher DBW than plane waves in an isotropic homogeneous medium (lower degree of TE-TM degeneracy) and lower LD than guided waves in a dielectric plate.

In this study, to reveal the TE-TM degeneracy conditions in dielectric metasurfaces, we calculate the dispersion characteristics of modes and their field distributions using the block-iterative frequency-domain method based on the open-source MIT Photonic Bands (MPB) package~\cite{johnson2001block}. The designs of metasurfaces are optimized with a tolerance of $0.001a$, which is close to or beyond the limit of fabrication precision.

\subsection{Operational Principle}

The electromagnetic response of dielectric metasurfaces is based on the resonant behavior of their constitutive particles, which may be controlled by choosing and tuning the corresponding modes of each resonator. It is similar to the unidirectional excitation by the magnetic circular dipoles~\cite{rodriguez2013near,li2015all} and Huygens' metasurfaces~\cite{decker2015high,chong2016efficient}, although higher-order modes can also be utilized~\cite{Evlyukhin_acsphotonics_2022, Allayarov_OptExpress_2024}.

In particular, the operational principle of our metasurface is based on the spectral overlapping between two mutually orthogonal Mie-type magnetic dipole modes of disk-shaped resonators. They are the HE$_{11\ell}$ (horizontal magnetic dipole) and TE$_{01\ell}$ (vertical magnetic dipole)  modes of a disk \cite{Mongia_1994}  (in the subscripts of the mode abbreviations, the third index $\ell$ denotes the order of variation of fields along the disk's height). The resonant frequency of the HE$_{11\ell}$ mode depends on the height $h$ of the resonator, while that of the TE$_{01\ell}$ mode depends on the diameter $d$ of the disk. These modes of the disk can be effectively excited by the $y$- and $z$-oriented magnetic dipoles, resulting in the excitation of the TM ($E_z \neq 0$, $H_y \neq 0$) and TE ($H_z \neq 0$, $E_y \neq 0$) guided modes, respectively,  propagating along the $x$-axis through the overall system. We should note that the origin of the TE-TM degeneracy, caused by the near-field spectral overlapping of the magnetic dipoles located within a plane perpendicular to the metasurface, is unique and has not been studied before.

Moreover, by changing the diameter-to-height ratio, it is possible to bring modes of the resonator close in frequency realizing the generalized Kerker condition~\cite{babicheva2017resonant, Shamkhi_PhysRevLett_2019, Shamkhi_PhysRevMaterials_2019} and thus achieving the polarization degeneracy. However, the Kerker condition for a single resonator is insufficient to achieve the overlapping of the TE- and TM-guided modes in a dielectric metasurface~\cite{Asadulina_lpor_2023}. The reason is the strong spatial dispersion, i.e. the interaction between the resonators forming an array, which is dependent on the parameter $a$. Therefore, to achieve the broadband TE-TM degeneracy of the guided waves in the given metasurface, the corresponding near-field parametric problem must be solved.

\begin{figure*}[t]
\centering
\includegraphics[width=1.0\linewidth]{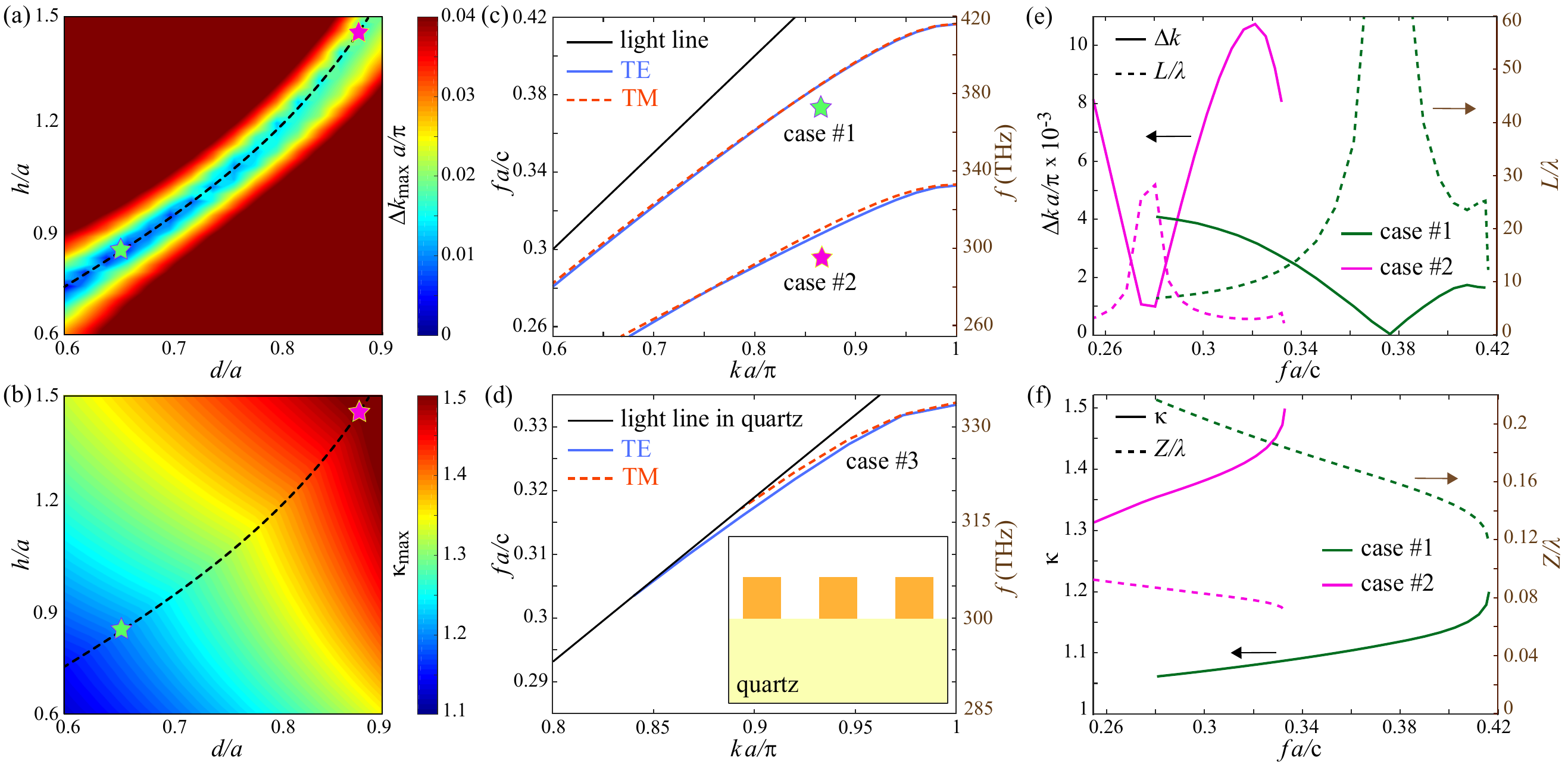}
\caption{Dependencies of the local maxima of (a) DBW $\Delta k_{max}$ and (b) LD $\kappa_{max}$ on the diameter-to-period $d/a$ and height-to-period $h/a$ ratios of the Si$_3$N$_4$ disks with $n = 2$. The dashed black line shows the optimal designs supporting the guided waves with the TE-TM degeneracy. The green and magenta stars correspond to the designs with minimal DBW and $\kappa = 1.2$ and $\kappa = 1.5$, respectively. (c,d) Dispersion characteristics of the TE (blue solid line) and TM (red dashed line) modes for (c) cases \#1 and \#2 marked by the stars in panels (a) and (b), and (d) case \#3 with a quartz ($n_\textrm{sub} = 1.45$) substrate according to Table~\ref{tab:tab_1}. The right bar in (c,d) corresponds to the frequency in THz for the metasurface with period $a = 300$~nm. The black line corresponds to the light line in (c) a vacuum and (d) quartz. Dependencies of (e) $\Delta k$ (solid lines) and $L_\lambda$ (dashed lines), and (f) $\kappa$ (solid lines) and $Z_\lambda$ (dashed lines) on the frequency when $\kappa_{max} = 1.2$ (green lines) and $\kappa_{max} = 1.5$ (magenta lines).}
\label{fig:fig_3}
\end{figure*}

\subsection{Targeted Metasurface Design}

The parameters of the problem are the refractive index of the disks, their diameter-to-height ratio, and the period of the structure. First, we analyze the best (i.e., with minimum DBW) designs for the metasurfaces at several specified LDs. The corresponding parametric dependencies are collected in Fig.~\ref{fig:fig_2}. For convenience, they are related to both the permittivity (lower scale) and refractive index (upper scale) of the disk material ($\varepsilon=n^2$). 

One can see that for different levels of LD, the dependence of the diameter-to-height ratio on the constitutive material parameter is typically the same. It linearly increases from $d/h \approx 0.64 \pm 0.12$ at $\varepsilon = 3.8$ ($n = 1.95$) up to $d/h \approx 1.12 \pm 0.03$ at $\varepsilon \approx 16 \pm 1$ ($n \approx 4 \pm 0.13$), and then smoothly decreases to $d/h \approx 1$ at $\varepsilon = 49$ ($n = 7$) for $\kappa$ ranging from 1.2 to 1.6 that corresponds to $L_\lambda$ values from 0.12 to 0.064, respectively [Fig.~\ref{fig:fig_2}(a)].  

The $h/a$ and $d/a$ are separately shown in Figs.~\ref{fig:fig_2}(b) and \ref{fig:fig_2}(c), respectively. The universal $d/a$ and $d/h$ dependencies averaged over different LD levels are shown in Fig.~\ref{fig:fig_2}(d). The obtained dependencies allow for a targeted selection of materials and principal geometric dimensions of the metasurface for its operation in the required spectral range (the corresponding ranges and suitable materials for the metasurface fabrication are shown as colored areas).

\subsection{Si$_3$N$_4$ Metasurface in Near-Infrared}

\begin{table}[b]
\centering
\caption{\textbf{Geometric and material parameters of several designs of a dielectric metasurface and corresponding maximum values of the LD and DBW within the specified wavelength range.}}
\begin{tabular}{@{}lccc@{}}
\toprule
Parameters & {\hspace{12pt} Case~\#1} & {\hspace{12pt} Case~\#2} & {\hspace{12pt} Case~\#3} \\
\hline
wavelength range & \multicolumn{3}{c}{\hspace{12pt} near-infrared (500--1700~nm)} \\
constitutive material & \multicolumn{3}{c}{\hspace{12pt} silicon nitride~\cite{beliaev2022optical}} \\
$n_d$ & \multicolumn{3}{c}{\hspace{12pt} 2.0} \\
$\varepsilon_d$ & \multicolumn{3}{c}{\hspace{12pt} 4.0} \\
substrate & \multicolumn{2}{c}{\hspace{12pt} absent} & {\hspace{12pt} quartz} \\
$d/a$ & {\hspace{12pt} 0.655} & {\hspace{12pt} 0.880} & {\hspace{12pt} 0.888} \\
$h/a$ & {\hspace{12pt} 0.823} & {\hspace{12pt} 1.456} & {\hspace{12pt} 0.957} \\
$\kappa_{max}$ & {\hspace{12pt} 1.2} & {\hspace{12pt} 1.5} & {\hspace{12pt} 1.04} \\
$Z_{min}/\lambda$ & {\hspace{12pt} 0.12} & {\hspace{12pt} 0.07} & {\hspace{12pt} 0.07} \\
$\Delta k_{max} \,(\pi/a) \times 10^{-3}$ & {\hspace{12pt} 4.1} & {\hspace{12pt} 10.7} & {\hspace{12pt} 4.1} \\
$L_{max}/\lambda$ & {\hspace{12pt} 898} & {\hspace{12pt} 28} & {\hspace{12pt} 50} \\
\bottomrule
\end{tabular}
\label{tab:tab_1}
\end{table}

As an illustrative example, we perform our numerical calculations for the metasurface made of Si$_3$N$_4$ [purple area in Fig.~\ref{fig:fig_2}(a)]. Such a metasurface can be a platform for CMOS-compatible optical elements~\cite{colburn2018broadband}, color meta-pixels~\cite{yang2020structural}, spatial modulators~\cite{sun2021electro}, and on-fiber light collectors~\cite{yermakov2020nanostructure} and sensors~\cite{Subramanian_PhotonRes_2015}. 
For reference, the set of material and geometrical parameters of the optical metasurfaces studied in this paper is summarized in Table~\ref{tab:tab_1}. Note, case \#3 is a metasurface deposited on a substrate. 

In our parametric study, we sweep the diameter and height of the silicon-nitride disks from $0.6a$ to $0.9a$ and $1.5a$, respectively, and calculate the related local maxima of the DBW [Fig.~\ref{fig:fig_3}(a)] and LD [Fig.~\ref{fig:fig_3}(b)] for each metasurface design. The black dashed lines show the optimal designs of the metasurface supporting the guided modes with the TE-TM degeneracy. Along this line, the degree of the TE-TM degeneracy changes from $\Delta k_{max} \approx 0.002\pi/a$ at $\kappa_{max} = 1.15$ to $\Delta k_{max} \approx 0.02\pi/a$ at $\kappa_{max} = 1.52$. The typical dispersion curves of the TE and TM modes when $\kappa_{max} = 1.2$ and $\kappa_{max} = 1.5$ corresponding to cases \#1 ($d/a = 0.655$, $h/a = 0.843$) and \#2 ($d/a = 0.88$, $h/a = 1.456$) from Table~\ref{tab:tab_1} are shown in Fig.~\ref{fig:fig_3}(c). At the fixed period of the metasurface ($a = 300$~nm), the resonant frequencies at the edge of the Brillouin zone ($k = \pi/a$, $\kappa = \kappa_{max}$) are 417~THz and 333~THz for the cases $\kappa_{max} = 1.2$ and $\kappa_{max} = 1.5$, respectively [Fig.~\ref{fig:fig_3}(c)]. 

The related dependencies of $\Delta k$ and $L_\lambda$, $\kappa$ and $Z_\lambda$ for these two cases are shown in Figs.~\ref{fig:fig_3}(e) and \ref{fig:fig_3}(f), respectively. One can notice that the DBW does not exceed $4.1\times10^{-3}\,\pi/a$ and $10.7\times10^{-3}\,\pi/a$ for the cases \#1 and \#2, respectively, while $\Delta k < 2\times10^{-3}\,\pi/a$ ($L_\lambda > 17$) in the frequency range $0.345c/a < f < 0.415c/a$ for $\kappa_{max} = 1.2$ and $\Delta k < 5\times10^{-3}\,\pi/a$ ($L_\lambda > 5.3$)  in the frequency range $0.264c/a < f < 0.293c/a$ for $\kappa_{max} = 1.5$. The distance $L$ saving the polarization degeneracy may achieve $200 \lambda$ and $25 \lambda$ in the vicinity of $f = 0.376 c/a$ and $f = 0.28 c/a$ for the cases \#1 and \#2, respectively. The average values of the DBW and LD in the wavenumber range $0.7$, $\pi/a < k < \pi/a$ are $1.6\times10^{-3}\,\pi/a$ and $0.16 \lambda$, $7.2\times10^{-3}\,\pi/a$ and $0.08 \lambda$ when $\kappa_{max} = 1.2$ and $\kappa_{max} = 1.5$, respectively.

\begin{figure*}[t]
\centering
\includegraphics[width=0.9\linewidth]{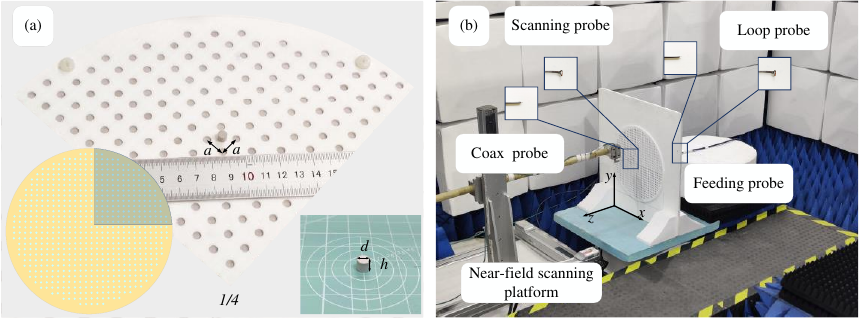}
\caption{(a) Schematic view and photographs of a fragment of the metasurface and individual ceramic resonator referenced to the rule. (b) Photograph of the near-field measurement setup, where pictures of the hand-made probes are given in the insets.}
\label{fig:fig_4}
\end{figure*}

\subsection{Impact of Substrate}

We have previously demonstrated several designs based on dielectric disks in a vacuum. From a practical point of view, the dielectric substrate should be taken into account. The impact of substrate can be compensated in the cases of low permittivity and/or small thickness of substrate via the slight enlargement of disks and decreasing the LD. The high index contrast between the resonators and substrate is inversely proportional to the effect of the substrate. To support this statement, we show in Fig.~\ref{fig:fig_3}(d) the dispersion characteristics of the metasurface ($d/a = 0.888$, $h/a = 0.957$) made of Si$_3$N$_4$ disks deposited on a quartz ($n_\textrm{sub} = 1.45$~\cite{malitson1965interspecimen}) substrate. This corresponds to case \#3 listed in Table~\ref{tab:tab_1}.

In this case, the DBW does not exceed $\Delta k < 4.1\times10^{-3} \,\pi/a$. The values of $L$ are typically tens of wavelengths exceeding $50\lambda$ near the corresponding frequency $0.332c/a$. The typical decay length is about tenths of a wavelength. So, the presence of a substrate made of a low-refractive-index material does not prevent the achievement of the TE-TM degeneracy in the metasurfaces under consideration.

\section{Experimental Verification \label{section:experimental}}

\subsection{Sample and Measurement Setup}

Although dielectric metasurfaces are mainly aimed at working in the optical spectral range, using quasi-optical (microwave) methods for their study is a widespread practice \cite{Filonov_ApplPhysLett_2012, Xu_AdvOptMater_2019, Yermakov_PhysRevX.11.031038,Kupriianov_JApplPhys_2023}. In this case, optical nanostructures are replaced by their prototypes, which are manufactured on a millimeter or centimeter scale. Most often, such structures are made of resonators from microwave ceramics~\cite{materials_2017}. Using such prototyping allows for a relatively quick and inexpensive proof of principle. Therefore, in this section, our goal is to demonstrate the TE-TM degeneracy of guided modes in a ceramic-based metasurface.

Given the constraints and capabilities of our experimental means, we have selected a microwave frequency band spanning from 8 to 11~GHz for our comprehensive analysis. The sample is assembled from two components: disk-shaped ceramic resonators and a holder. Figure~\ref{fig:fig_4}(a) illustrates a schematic representation of a metasurface structure, photographs of a quarter-scale sample, and individual ceramic resonator. The resonators are fabricated using polymer-based Taizhou Wangling TP-1/2 ceramics, a high-permittivity material. As per the manufacturer's datasheet \cite{wangling}, this material exhibits a relative permittivity of $\varepsilon = 25$, and a dissipation factor $\text{tan} \delta = 0.002$ referenced to the frequency of 10 GHz. This material is available on the market in the form of ceramic plates with fixed thicknesses. For our study, we chose plates with a thickness of 5 mm. From these plates, the resonators are made by a water-jet cutting technique.

Thus, the height of the resonators is determined by the thickness of the plates ($h=5$~mm), whereas the disks' diameter is obtained from the solution of the optimization problem. The period of the structure is also a subject of optimization for $\kappa_{max} = 1.25$. The diameter of the disks and period of the structure obtained are $d = 5.32$~mm and $a = 12.5$~mm, respectively, i.e. $d/a = 0.426$ and $h/a = 0.4$.

\begin{figure*}[t]
\centering
\includegraphics[width=0.9\linewidth]{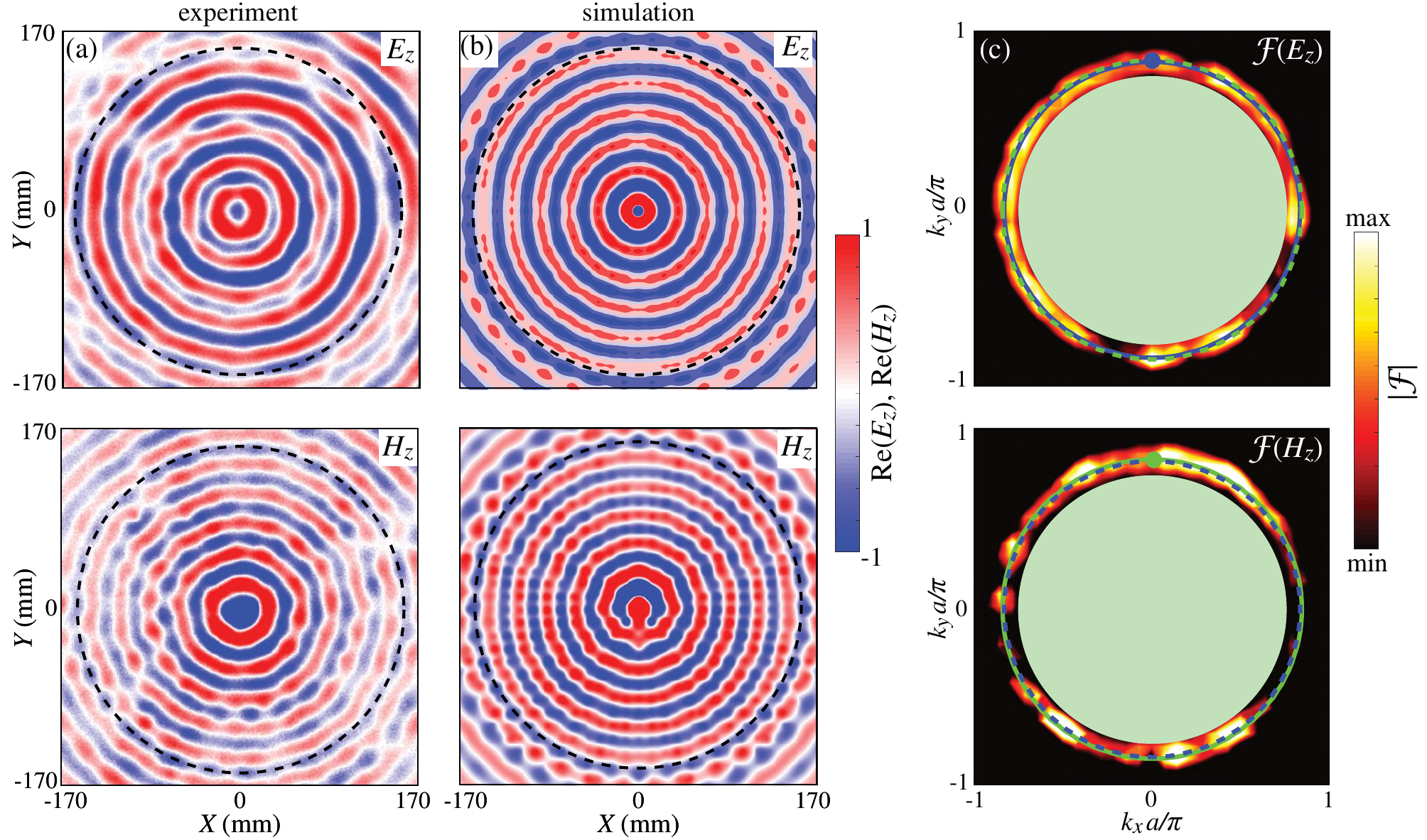}
\caption{(a) Experimentally measured and (b) numerically simulated spatial distributions of the real part (normalized on the maximal value) of the normal component of electric field (upper row) and magnetic field (bottom row) excited by $z$-oriented probe and loop in the $x$-$y$ plane, respectively, and (c) their Fourier spectrum within the first Brillouin zone where the maximal values correspond to the isofrequency contours. The black dashed lines show the edge of the metasurface. The light green circles outline the light line in a vacuum. The dashed lines and dots correspond to the isofrequency contours calculated numerically and the dispersion points along $y$-direction extracted from the experimental data (blue - TM, green - TE).}
\label{fig:fig_5}
\end{figure*}

In our experiments, a rigid ROHACELL-HF71 foam plate, possessing air-like dielectric properties with a relative permittivity of $\varepsilon_\textrm{sub} = 1.09$, serves as a holder. This plate, measuring 20 mm in thickness, is precision-milled to accommodate resonators in a periodic array with the given lattice period. The plate's holes are tailored to match the resonators' diameter and height. In the final sample, the grating holder is ingeniously designed as an assembly of four quarter-circular segments, which are integrated to form a circular configuration. The resulting metasurface comprises 641 resonators, arranged in a periodic pattern forming a circle with a radius of 190~mm.

For the near-field study, we employ a LINBOU NF-3 three-axis scanner platform, depicted in Fig. \ref{fig:fig_4}(b). This platform features three linear rails anchored to a robust metallic frame, establishing a standard right-handed Cartesian coordinate system. The horizontal linear motion, confined to the $x$-$z$ plane, is facilitated by two of the rails. The third rail, rigidly affixed to its counterparts, supports a wooden arm that cradles a near-field probe, enabling vertical linear motion along the $y$-axis. The movement across the $x$, $y$, and $z$ axes allows the probe to delineate a precise trajectory above the metasurface. The sample is positioned upright on a stable measuring table, anchored and stabilized by four support bases. During the measurements, a stationary feeding probe was centrally located within the structure, positioned 5 mm from the sample. Concurrently, the scanning probe is precisely positioned by a near-field scanner to a plane parallel to the metasurface, on the opposite side of the structure, with a meticulously maintained gap of 5 mm between the probe tip and the sample. This scanning probe is connected to a vector network analyzer (VNA) Rohde \& Schwarz ZVA-50.

To perform near-field scanning, two types of self-made probes are used. For the measurement of the TM-guided modes, electric (coaxial) monopoles were used both as a source and as a probe to measure the normal component of the electric field. Similarly, for the measurement of the TE-guided modes, magnetic (loop) dipoles are utilized in a dual capacity, as both the source and the probe, to accurately measure the normal component of the magnetic field. For the near-field measurements, the probe moves in the $x$-$y$ plane over the scanning area $340 \times 340$ mm$^2$ with a step of 1~mm. The scanner linear positioning data are obtained from stepper motors with encoders. At each sampled location, the amplitude distributions of the $z$-component of the field, whether electric or magnetic, are meticulously collected across the operational frequency spectrum, with a frequency step of 0.01~GHz.

\subsection{Measurement Results}

We use independently the excitation probe oriented along the $z$-axis orthogonally to the metasurface and loop lying within the $x$-$y$ plane parallel to the metasurface located in the center of the structure at the height of 5~mm from the top of the disks. These probes and loop act as the vertical electric and magnetic dipoles exciting TM ($E_z \neq 0$, $H_y \neq 0$) and TE ($H_z \neq 0$, $E_y \neq 0$) guided modes, respectively. We use the same probe and loop to measure the normal components of the electric and magnetic fields. As a result, we measure the spatial distributions of the $E_z$ and $H_z$ field components at 9.5~GHz related to the TM and TE modes, respectively [Fig.~\ref{fig:fig_5}(a)], that coincides well with the corresponding numerical calculations [Fig.~\ref{fig:fig_5}(b)]. Both modes within the unit cell satisfy condition $k_s = k_p \approx \pi/a$. Then, we extract the isofrequency contours by applying the two-dimensional Fourier transform to the calculated distribution of the normal components of the electromagnetic fields [Fig.~\ref{fig:fig_5}(c)] following the standard procedure~\cite{dockrey2016direct, yang2017hyperbolic, yermakov2018experimental}. On the other hand, we calculate the corresponding isofrequency contour numerically. We claim a good coincidence between the numerical and experimental results and between the TE and TM polarizations for the isofrequency contours signifying the TE-TM degeneracy in all directions within a plane of metasurface. 

\begin{figure}[t]
\centering
\includegraphics[width=1.0\linewidth]{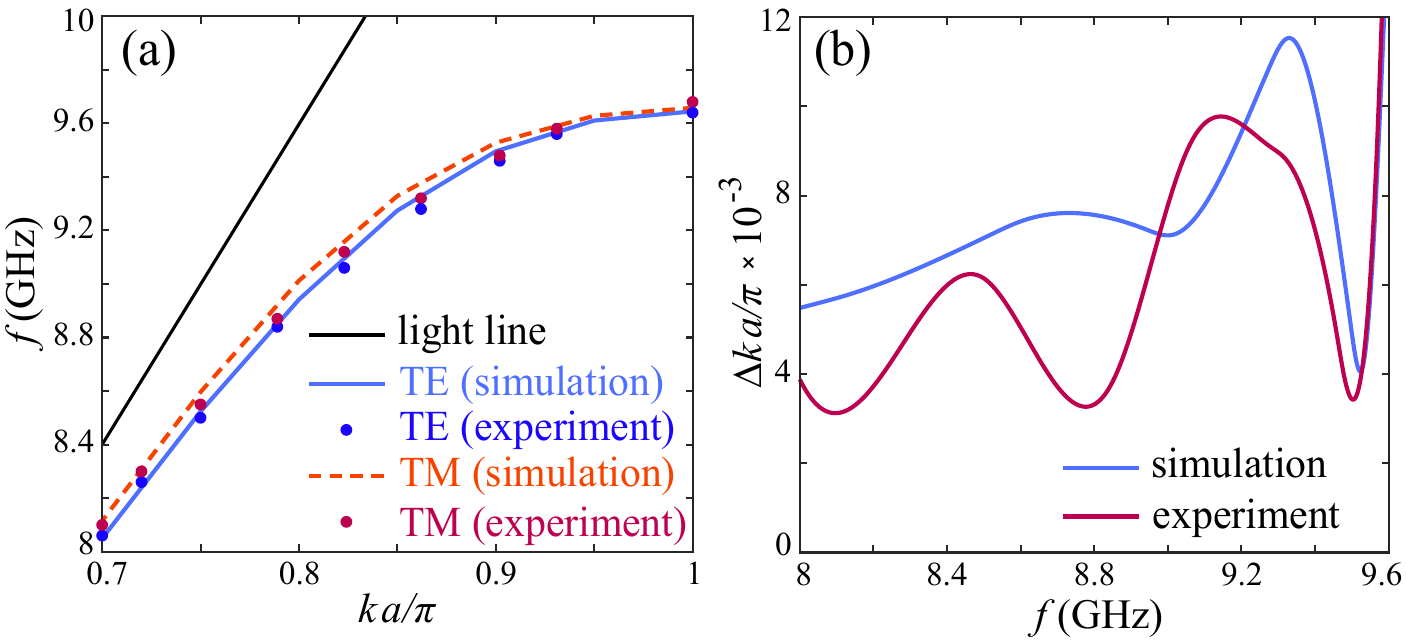}
\caption{(a) Dispersion characteristics of the TE (blue) and TM (red) guided modes propagated along dielectric metasurface supporting the TE-TM degeneracy. The lines and dots correspond to the simulated and measured data, respectively. The black line shows the light line in a vacuum. (b) Dependence of $\Delta k$ on the frequency for the metasurface under consideration retrieved numerically (blue line) and experimentally (red line). }
\label{fig:fig_6}
\end{figure}

Finally, we extract the dispersion of the TE- and TM-guided modes along the $x$- or $y$-direction from the experimentally measured field distributions and compare them with corresponding numerical calculations [Fig.~\ref{fig:fig_6}(a)]. The related DBW does not exceed $\Delta k < 12\times10^{-3} \, \pi/a$ in the frequency range up to 9.6~GHz for both numerical and experimental data [Fig.~\ref{fig:fig_6}(b)]. Even more, one can notice the same behavior of DBW in numerical and experimental results such as dips and peaks in $\Delta k (f)$. So, this microwave experiment confirms the relevance of the introduced design approach and presents the first implementation of the metasurface with the TE-TM degenerate-guided modes.

\section{Conclusions}

In this study, we proposed and experimentally demonstrated a dielectric metasurface capable of controlling the polarization of highly localized guided waves. The operational principle is based on the overlapping Mie-type magnetic dipole modes of disk-shaped resonators forming the metasurface. Thus, the guided waves in such a metasurface simultaneously achieve both the polarization degree of freedom and a high degree of localization. Furthermore, we have utilized several specific designs of the metasurface, made from silicon nitride disks and operated in the near-infrared spectrum, to demonstrate the universality of our principle. In a similar vein, we have also established that dielectric metasurface, whether mounted on a quartz substrate or free-standing, can indeed exhibit TE-TM mode degeneracy.

Experimental validation in the microwave frequency range confirms our principle that guided modes with high polarization degeneracy can be propagated in all in-plane directions within a metasurface. Moreover, this microwave experiment represents the inaugural implementation of a metasurface featuring TE-TM degenerate-guided modes. The metasurface achieves a simple structure, a high degree of localization, and a large degree of polarization freedom simultaneously. 

The wide application of nanostructured dielectric resonators defines the direction of all-dielectric nanophotonics~\cite{Baranov_Optica_2017} and Mie-tronics~\cite{won2019into,kivshar2022rise}. The interference between Mie-like dipole and higher-order multipole resonances opens up new possibilities for light manipulation, directional scattering, nonlinear optics, sensing, and many other areas~\cite{koshelev2020dielectric,koshelev2020subwavelength}. The results of this work complement the previous study discovering a new type of interaction between Mie resonances leading to the near-field polarization degree of freedom. It opens up new opportunities for the all-dielectric nanophotonic platform. 

We believe that our findings can lay the groundwork for future research on optical platforms supporting localized guided waves with TE-TM degeneracy. Furthermore, the designed metasurface could be extended to a novel platform for planar photonic devices with polarization capabilities, thereby opening up plenty of applications, especially for polarization converters and optical sensors.

\section*{Acknowledgements}
S.P. and V.T. acknowledge funding from the European Union's Horizon 2020 Research and Innovation programme under Grant Agreement no.~871072. O.Y. acknowledges the support of the Alexander von Humboldt Foundation within the framework of the Humboldt Research Fellowship for postdoctoral researchers and the IRTAP-Ukraine program of APS supporting physicists impacted by the Russian invasion of Ukraine.

\bibliography{references}

\begin{thebibliography}{60}%
\makeatletter
\providecommand \@ifxundefined [1]{%
 \@ifx{#1\undefined}
}%
\providecommand \@ifnum [1]{%
 \ifnum #1\expandafter \@firstoftwo
 \else \expandafter \@secondoftwo
 \fi
}%
\providecommand \@ifx [1]{%
 \ifx #1\expandafter \@firstoftwo
 \else \expandafter \@secondoftwo
 \fi
}%
\providecommand \natexlab [1]{#1}%
\providecommand \enquote  [1]{``#1''}%
\providecommand \bibnamefont  [1]{#1}%
\providecommand \bibfnamefont [1]{#1}%
\providecommand \citenamefont [1]{#1}%
\providecommand \href@noop [0]{\@secondoftwo}%
\providecommand \href [0]{\begingroup \@sanitize@url \@href}%
\providecommand \@href[1]{\@@startlink{#1}\@@href}%
\providecommand \@@href[1]{\endgroup#1\@@endlink}%
\providecommand \@sanitize@url [0]{\catcode `\\12\catcode `\$12\catcode `\&12\catcode `\#12\catcode `\^12\catcode `\_12\catcode `\%12\relax}%
\providecommand \@@startlink[1]{}%
\providecommand \@@endlink[0]{}%
\providecommand \url  [0]{\begingroup\@sanitize@url \@url }%
\providecommand \@url [1]{\endgroup\@href {#1}{\urlprefix }}%
\providecommand \urlprefix  [0]{URL }%
\providecommand \Eprint [0]{\href }%
\providecommand \doibase [0]{https://doi.org/}%
\providecommand \selectlanguage [0]{\@gobble}%
\providecommand \bibinfo  [0]{\@secondoftwo}%
\providecommand \bibfield  [0]{\@secondoftwo}%
\providecommand \translation [1]{[#1]}%
\providecommand \BibitemOpen [0]{}%
\providecommand \bibitemStop [0]{}%
\providecommand \bibitemNoStop [0]{.\EOS\space}%
\providecommand \EOS [0]{\spacefactor3000\relax}%
\providecommand \BibitemShut  [1]{\csname bibitem#1\endcsname}%
\let\auto@bib@innerbib\@empty
\bibitem [{\citenamefont {Born}\ and\ \citenamefont {Wolf}(1959)}]{born2013principles}%
  \BibitemOpen
  \bibfield  {author} {\bibinfo {author} {\bibfnamefont {M.}~\bibnamefont {Born}}\ and\ \bibinfo {author} {\bibfnamefont {E.}~\bibnamefont {Wolf}},\ }\href@noop {} {\emph {\bibinfo {title} {Principles of Optics: Electromagnetic Theory of Propagation, Interference and Diffraction of Light}}}\ (\bibinfo  {publisher} {Pergamon Press Ltd.},\ \bibinfo {address} {New York, NY},\ \bibinfo {year} {1959})\BibitemShut {NoStop}%
\bibitem [{\citenamefont {Mu}\ \emph {et~al.}(2019)\citenamefont {Mu}, \citenamefont {Fan}, \citenamefont {Chen}, \citenamefont {Xu}, \citenamefont {Xiong}, \citenamefont {Zhang}, \citenamefont {Wang},\ and\ \citenamefont {Chang}}]{Mu_PhotonRes_2019}%
  \BibitemOpen
  \bibfield  {author} {\bibinfo {author} {\bibfnamefont {Q.}~\bibnamefont {Mu}}, \bibinfo {author} {\bibfnamefont {F.}~\bibnamefont {Fan}}, \bibinfo {author} {\bibfnamefont {S.}~\bibnamefont {Chen}}, \bibinfo {author} {\bibfnamefont {S.}~\bibnamefont {Xu}}, \bibinfo {author} {\bibfnamefont {C.}~\bibnamefont {Xiong}}, \bibinfo {author} {\bibfnamefont {X.}~\bibnamefont {Zhang}}, \bibinfo {author} {\bibfnamefont {X.}~\bibnamefont {Wang}},\ and\ \bibinfo {author} {\bibfnamefont {S.}~\bibnamefont {Chang}},\ }\bibfield  {title} {\bibinfo {title} {Tunable magneto-optical polarization device for terahertz waves based on {InSb} and its plasmonic structure},\ }\href {https://doi.org/10.1364/PRJ.7.000325} {\bibfield  {journal} {\bibinfo  {journal} {Photon. Res.}\ }\textbf {\bibinfo {volume} {7}},\ \bibinfo {pages} {325} (\bibinfo {year} {2019})}\BibitemShut {NoStop}%
\bibitem [{\citenamefont {Liu}\ \emph {et~al.}(2024{\natexlab{a}})\citenamefont {Liu}, \citenamefont {Zhang}, \citenamefont {Liu}, \citenamefont {Yu}, \citenamefont {Wu}, \citenamefont {Xiao}, \citenamefont {Huang},\ and\ \citenamefont {Miroshnichenko}}]{Xiao_PhysRevApplied_2024}%
  \BibitemOpen
  \bibfield  {author} {\bibinfo {author} {\bibfnamefont {T.}~\bibnamefont {Liu}}, \bibinfo {author} {\bibfnamefont {D.}~\bibnamefont {Zhang}}, \bibinfo {author} {\bibfnamefont {W.}~\bibnamefont {Liu}}, \bibinfo {author} {\bibfnamefont {T.}~\bibnamefont {Yu}}, \bibinfo {author} {\bibfnamefont {F.}~\bibnamefont {Wu}}, \bibinfo {author} {\bibfnamefont {S.}~\bibnamefont {Xiao}}, \bibinfo {author} {\bibfnamefont {L.}~\bibnamefont {Huang}},\ and\ \bibinfo {author} {\bibfnamefont {A.~E.}\ \bibnamefont {Miroshnichenko}},\ }\bibfield  {title} {\bibinfo {title} {Phase-change nonlocal metasurfaces for dynamic wave-front manipulation},\ }\href {https://doi.org/10.1103/PhysRevApplied.21.044004} {\bibfield  {journal} {\bibinfo  {journal} {Phys. Rev. Appl.}\ }\textbf {\bibinfo {volume} {21}},\ \bibinfo {pages} {044004} (\bibinfo {year} {2024}{\natexlab{a}})}\BibitemShut {NoStop}%
\bibitem [{\citenamefont {Zhu}\ \emph {et~al.}(2024)\citenamefont {Zhu}, \citenamefont {Li}, \citenamefont {Qin}, \citenamefont {Jiang}, \citenamefont {Wang}, \citenamefont {Chen}, \citenamefont {Wang}, \citenamefont {Pang},\ and\ \citenamefont {Qu}}]{Zhu_PhotonRes_2024}%
  \BibitemOpen
  \bibfield  {author} {\bibinfo {author} {\bibfnamefont {Z.}~\bibnamefont {Zhu}}, \bibinfo {author} {\bibfnamefont {Y.}~\bibnamefont {Li}}, \bibinfo {author} {\bibfnamefont {Z.}~\bibnamefont {Qin}}, \bibinfo {author} {\bibfnamefont {L.}~\bibnamefont {Jiang}}, \bibinfo {author} {\bibfnamefont {W.}~\bibnamefont {Wang}}, \bibinfo {author} {\bibfnamefont {H.}~\bibnamefont {Chen}}, \bibinfo {author} {\bibfnamefont {J.}~\bibnamefont {Wang}}, \bibinfo {author} {\bibfnamefont {Y.}~\bibnamefont {Pang}},\ and\ \bibinfo {author} {\bibfnamefont {S.}~\bibnamefont {Qu}},\ }\bibfield  {title} {\bibinfo {title} {Miura origami based reconfigurable polarization converter for multifunctional control of electromagnetic waves},\ }\href {https://doi.org/10.1364/PRJ.504027} {\bibfield  {journal} {\bibinfo  {journal} {Photon. Res.}\ }\textbf {\bibinfo {volume} {12}},\ \bibinfo {pages} {581} (\bibinfo {year} {2024})}\BibitemShut {NoStop}%
\bibitem [{\citenamefont {Collett}(2005)}]{collett2005field}%
  \BibitemOpen
  \bibfield  {author} {\bibinfo {author} {\bibfnamefont {E.}~\bibnamefont {Collett}},\ }\href@noop {} {\emph {\bibinfo {title} {Field Guide to Polarization}}},\ edited by\ \bibinfo {editor} {\bibfnamefont {J.~E.}\ \bibnamefont {Greivenkamp}},\ \bibinfo {series} {SPIE Field Guides}, Vol.\ \bibinfo {volume} {FG05}\ (\bibinfo  {publisher} {SPIE Press},\ \bibinfo {address} {Bellingham, WA},\ \bibinfo {year} {2005})\BibitemShut {NoStop}%
\bibitem [{\citenamefont {Chavez~Boggio}\ \emph {et~al.}(2014)\citenamefont {Chavez~Boggio}, \citenamefont {Bodenm\"{u}ller}, \citenamefont {Fremberg}, \citenamefont {Haynes}, \citenamefont {Roth}, \citenamefont {Eisermann}, \citenamefont {Lisker}, \citenamefont {Zimmermann},\ and\ \citenamefont {B\"{o}hm}}]{ChavezBoggio_JOSAB_2014}%
  \BibitemOpen
  \bibfield  {author} {\bibinfo {author} {\bibfnamefont {J.~M.}\ \bibnamefont {Chavez~Boggio}}, \bibinfo {author} {\bibfnamefont {D.}~\bibnamefont {Bodenm\"{u}ller}}, \bibinfo {author} {\bibfnamefont {T.}~\bibnamefont {Fremberg}}, \bibinfo {author} {\bibfnamefont {R.}~\bibnamefont {Haynes}}, \bibinfo {author} {\bibfnamefont {M.~M.}\ \bibnamefont {Roth}}, \bibinfo {author} {\bibfnamefont {R.}~\bibnamefont {Eisermann}}, \bibinfo {author} {\bibfnamefont {M.}~\bibnamefont {Lisker}}, \bibinfo {author} {\bibfnamefont {L.}~\bibnamefont {Zimmermann}},\ and\ \bibinfo {author} {\bibfnamefont {M.}~\bibnamefont {B\"{o}hm}},\ }\bibfield  {title} {\bibinfo {title} {Dispersion engineered silicon nitride waveguides by geometrical and refractive-index optimization},\ }\href {https://doi.org/10.1364/JOSAB.31.002846} {\bibfield  {journal} {\bibinfo  {journal} {J. Opt. Soc. Am. B}\ }\textbf {\bibinfo {volume} {31}},\ \bibinfo {pages} {2846} (\bibinfo {year} {2014})}\BibitemShut {NoStop}%
\bibitem [{\citenamefont {Zhao}\ \emph {et~al.}(2024)\citenamefont {Zhao}, \citenamefont {Peng}, \citenamefont {Zhu}, \citenamefont {Liu}, \citenamefont {Hu}, \citenamefont {Shi},\ and\ \citenamefont {Dai}}]{Zhao_PhotonRes_2024}%
  \BibitemOpen
  \bibfield  {author} {\bibinfo {author} {\bibfnamefont {W.}~\bibnamefont {Zhao}}, \bibinfo {author} {\bibfnamefont {Y.}~\bibnamefont {Peng}}, \bibinfo {author} {\bibfnamefont {M.}~\bibnamefont {Zhu}}, \bibinfo {author} {\bibfnamefont {R.}~\bibnamefont {Liu}}, \bibinfo {author} {\bibfnamefont {X.}~\bibnamefont {Hu}}, \bibinfo {author} {\bibfnamefont {Y.}~\bibnamefont {Shi}},\ and\ \bibinfo {author} {\bibfnamefont {D.}~\bibnamefont {Dai}},\ }\bibfield  {title} {\bibinfo {title} {Ultracompact silicon on-chip polarization controller},\ }\href {https://doi.org/10.1364/PRJ.499801} {\bibfield  {journal} {\bibinfo  {journal} {Photon. Res.}\ }\textbf {\bibinfo {volume} {12}},\ \bibinfo {pages} {183} (\bibinfo {year} {2024})}\BibitemShut {NoStop}%
\bibitem [{\citenamefont {Fesenko}\ \emph {et~al.}(2016)\citenamefont {Fesenko}, \citenamefont {Tuz}, \citenamefont {Shulika},\ and\ \citenamefont {Sukhoivanov}}]{Fesenko_Nanophot_2016}%
  \BibitemOpen
  \bibfield  {author} {\bibinfo {author} {\bibfnamefont {V.~I.}\ \bibnamefont {Fesenko}}, \bibinfo {author} {\bibfnamefont {V.~R.}\ \bibnamefont {Tuz}}, \bibinfo {author} {\bibfnamefont {O.~V.}\ \bibnamefont {Shulika}},\ and\ \bibinfo {author} {\bibfnamefont {I.~A.}\ \bibnamefont {Sukhoivanov}},\ }\bibfield  {title} {\bibinfo {title} {Dispersion properties of {Kolakoski}-cladding hollow-core nanophotonic {Bragg} waveguide},\ }\href {https://doi.org/doi:10.1515/nanoph-2016-0025} {\bibfield  {journal} {\bibinfo  {journal} {Nanophotonics}\ }\textbf {\bibinfo {volume} {5}},\ \bibinfo {pages} {556} (\bibinfo {year} {2016})}\BibitemShut {NoStop}%
\bibitem [{\citenamefont {Fesenko}\ and\ \citenamefont {Tuz}(2016)}]{Fesenko_OptCommun_2016}%
  \BibitemOpen
  \bibfield  {author} {\bibinfo {author} {\bibfnamefont {V.~I.}\ \bibnamefont {Fesenko}}\ and\ \bibinfo {author} {\bibfnamefont {V.~R.}\ \bibnamefont {Tuz}},\ }\bibfield  {title} {\bibinfo {title} {Dispersion blue-shift in an aperiodic {Bragg} reflection waveguide},\ }\href {https://doi.org/https://doi.org/10.1016/j.optcom.2015.12.016} {\bibfield  {journal} {\bibinfo  {journal} {Opt. Commun.}\ }\textbf {\bibinfo {volume} {365}},\ \bibinfo {pages} {225} (\bibinfo {year} {2016})}\BibitemShut {NoStop}%
\bibitem [{\citenamefont {Benedikovic}\ \emph {et~al.}(2017)\citenamefont {Benedikovic}, \citenamefont {Berciano}, \citenamefont {Alonso-Ramos}, \citenamefont {Le~Roux}, \citenamefont {Cassan}, \citenamefont {Marris-Morini},\ and\ \citenamefont {Vivien}}]{Benedikovic_OptExpress_2017}%
  \BibitemOpen
  \bibfield  {author} {\bibinfo {author} {\bibfnamefont {D.}~\bibnamefont {Benedikovic}}, \bibinfo {author} {\bibfnamefont {M.}~\bibnamefont {Berciano}}, \bibinfo {author} {\bibfnamefont {C.}~\bibnamefont {Alonso-Ramos}}, \bibinfo {author} {\bibfnamefont {X.}~\bibnamefont {Le~Roux}}, \bibinfo {author} {\bibfnamefont {E.}~\bibnamefont {Cassan}}, \bibinfo {author} {\bibfnamefont {D.}~\bibnamefont {Marris-Morini}},\ and\ \bibinfo {author} {\bibfnamefont {L.}~\bibnamefont {Vivien}},\ }\bibfield  {title} {\bibinfo {title} {Dispersion control of silicon nanophotonic waveguides using sub-wavelength grating metamaterials in near- and mid-{IR} wavelengths},\ }\href {https://doi.org/10.1364/OE.25.019468} {\bibfield  {journal} {\bibinfo  {journal} {Opt. Express}\ }\textbf {\bibinfo {volume} {25}},\ \bibinfo {pages} {19468} (\bibinfo {year} {2017})}\BibitemShut {NoStop}%
\bibitem [{\citenamefont {Teng}\ \emph {et~al.}(2019)\citenamefont {Teng}, \citenamefont {Zhang}, \citenamefont {Wang}, \citenamefont {Liu},\ and\ \citenamefont {Lv}}]{Teng_PhotonRes_2019}%
  \BibitemOpen
  \bibfield  {author} {\bibinfo {author} {\bibfnamefont {S.}~\bibnamefont {Teng}}, \bibinfo {author} {\bibfnamefont {Q.}~\bibnamefont {Zhang}}, \bibinfo {author} {\bibfnamefont {H.}~\bibnamefont {Wang}}, \bibinfo {author} {\bibfnamefont {L.}~\bibnamefont {Liu}},\ and\ \bibinfo {author} {\bibfnamefont {H.}~\bibnamefont {Lv}},\ }\bibfield  {title} {\bibinfo {title} {Conversion between polarization states based on a metasurface},\ }\href {https://doi.org/10.1364/PRJ.7.000246} {\bibfield  {journal} {\bibinfo  {journal} {Photon. Res.}\ }\textbf {\bibinfo {volume} {7}},\ \bibinfo {pages} {246} (\bibinfo {year} {2019})}\BibitemShut {NoStop}%
\bibitem [{\citenamefont {Pu}\ \emph {et~al.}(2019)\citenamefont {Pu}, \citenamefont {Guo}, \citenamefont {Ma}, \citenamefont {Li},\ and\ \citenamefont {Luo}}]{Pu_AdvOptMater_2019}%
  \BibitemOpen
  \bibfield  {author} {\bibinfo {author} {\bibfnamefont {M.}~\bibnamefont {Pu}}, \bibinfo {author} {\bibfnamefont {Y.}~\bibnamefont {Guo}}, \bibinfo {author} {\bibfnamefont {X.}~\bibnamefont {Ma}}, \bibinfo {author} {\bibfnamefont {X.}~\bibnamefont {Li}},\ and\ \bibinfo {author} {\bibfnamefont {X.}~\bibnamefont {Luo}},\ }\bibfield  {title} {\bibinfo {title} {Methodologies for on-demand dispersion engineering of waves in metasurfaces},\ }\href {https://doi.org/https://doi.org/10.1002/adom.201801376} {\bibfield  {journal} {\bibinfo  {journal} {Adv. Opt. Mater.}\ }\textbf {\bibinfo {volume} {7}},\ \bibinfo {pages} {1801376} (\bibinfo {year} {2019})}\BibitemShut {NoStop}%
\bibitem [{\citenamefont {Liu}\ \emph {et~al.}(2024{\natexlab{b}})\citenamefont {Liu}, \citenamefont {Li},\ and\ \citenamefont {Xiao}}]{Xiao_PhysRevResearch_2024}%
  \BibitemOpen
  \bibfield  {author} {\bibinfo {author} {\bibfnamefont {T.}~\bibnamefont {Liu}}, \bibinfo {author} {\bibfnamefont {J.}~\bibnamefont {Li}},\ and\ \bibinfo {author} {\bibfnamefont {S.}~\bibnamefont {Xiao}},\ }\bibfield  {title} {\bibinfo {title} {Single-sized phase-change metasurfaces for dynamic information multiplexing and encryption},\ }\href {https://doi.org/10.1103/PhysRevResearch.6.023258} {\bibfield  {journal} {\bibinfo  {journal} {Phys. Rev. Res.}\ }\textbf {\bibinfo {volume} {6}},\ \bibinfo {pages} {023258} (\bibinfo {year} {2024}{\natexlab{b}})}\BibitemShut {NoStop}%
\bibitem [{\citenamefont {Garanovich}\ \emph {et~al.}(2012)\citenamefont {Garanovich}, \citenamefont {Longhi}, \citenamefont {Sukhorukov},\ and\ \citenamefont {Kivshar}}]{Garanovich_PhysRep_2012}%
  \BibitemOpen
  \bibfield  {author} {\bibinfo {author} {\bibfnamefont {I.~L.}\ \bibnamefont {Garanovich}}, \bibinfo {author} {\bibfnamefont {S.}~\bibnamefont {Longhi}}, \bibinfo {author} {\bibfnamefont {A.~A.}\ \bibnamefont {Sukhorukov}},\ and\ \bibinfo {author} {\bibfnamefont {Y.~S.}\ \bibnamefont {Kivshar}},\ }\bibfield  {title} {\bibinfo {title} {Light propagation and localization in modulated photonic lattices and waveguides},\ }\href {https://doi.org/https://doi.org/10.1016/j.physrep.2012.03.005} {\bibfield  {journal} {\bibinfo  {journal} {Phys. Rep.}\ }\textbf {\bibinfo {volume} {518}},\ \bibinfo {pages} {1} (\bibinfo {year} {2012})}\BibitemShut {NoStop}%
\bibitem [{\citenamefont {Halir}\ \emph {et~al.}(2015)\citenamefont {Halir}, \citenamefont {Bock}, \citenamefont {Cheben}, \citenamefont {Ortega-Mo{\~n}ux}, \citenamefont {Alonso-Ramos}, \citenamefont {Schmid}, \citenamefont {Lapointe}, \citenamefont {Xu}, \citenamefont {Wang{\"u}emert-P{\'e}rez}, \citenamefont {Molina-Fern{\'a}ndez},\ and\ \citenamefont {Janz}}]{Halir_lpor_2014}%
  \BibitemOpen
  \bibfield  {author} {\bibinfo {author} {\bibfnamefont {R.}~\bibnamefont {Halir}}, \bibinfo {author} {\bibfnamefont {P.~J.}\ \bibnamefont {Bock}}, \bibinfo {author} {\bibfnamefont {P.}~\bibnamefont {Cheben}}, \bibinfo {author} {\bibfnamefont {A.}~\bibnamefont {Ortega-Mo{\~n}ux}}, \bibinfo {author} {\bibfnamefont {C.}~\bibnamefont {Alonso-Ramos}}, \bibinfo {author} {\bibfnamefont {J.~H.}\ \bibnamefont {Schmid}}, \bibinfo {author} {\bibfnamefont {J.}~\bibnamefont {Lapointe}}, \bibinfo {author} {\bibfnamefont {D.-X.}\ \bibnamefont {Xu}}, \bibinfo {author} {\bibfnamefont {J.~G.}\ \bibnamefont {Wang{\"u}emert-P{\'e}rez}}, \bibinfo {author} {\bibfnamefont {{\'I}.}~\bibnamefont {Molina-Fern{\'a}ndez}},\ and\ \bibinfo {author} {\bibfnamefont {S.}~\bibnamefont {Janz}},\ }\bibfield  {title} {\bibinfo {title} {Waveguide sub-wavelength structures: a review of principles and applications},\ }\href {https://doi.org/https://doi.org/10.1002/lpor.201400083} {\bibfield  {journal} {\bibinfo  {journal} {Laser Photonics Rev.}\
  }\textbf {\bibinfo {volume} {9}},\ \bibinfo {pages} {25} (\bibinfo {year} {2015})}\BibitemShut {NoStop}%
\bibitem [{\citenamefont {Quaranta}\ \emph {et~al.}(2018)\citenamefont {Quaranta}, \citenamefont {Basset}, \citenamefont {Martin},\ and\ \citenamefont {Gallinet}}]{Quaranta_lpor_2018}%
  \BibitemOpen
  \bibfield  {author} {\bibinfo {author} {\bibfnamefont {G.}~\bibnamefont {Quaranta}}, \bibinfo {author} {\bibfnamefont {G.}~\bibnamefont {Basset}}, \bibinfo {author} {\bibfnamefont {O.~J.~F.}\ \bibnamefont {Martin}},\ and\ \bibinfo {author} {\bibfnamefont {B.}~\bibnamefont {Gallinet}},\ }\bibfield  {title} {\bibinfo {title} {Recent advances in resonant waveguide gratings},\ }\href {https://doi.org/https://doi.org/10.1002/lpor.201800017} {\bibfield  {journal} {\bibinfo  {journal} {Laser Photonics Rev.}\ }\textbf {\bibinfo {volume} {12}},\ \bibinfo {pages} {1800017} (\bibinfo {year} {2018})}\BibitemShut {NoStop}%
\bibitem [{\citenamefont {Meng}\ \emph {et~al.}(2021)\citenamefont {Meng}, \citenamefont {Chen}, \citenamefont {Lu}, \citenamefont {Ding}, \citenamefont {Cusano}, \citenamefont {Fan}, \citenamefont {Hu}, \citenamefont {Wang}, \citenamefont {Xie}, \citenamefont {Liu}, \citenamefont {Yang}, \citenamefont {Liu}, \citenamefont {Gong}, \citenamefont {Xiao}, \citenamefont {Sun}, \citenamefont {Zhang}, \citenamefont {Yuan},\ and\ \citenamefont {Ni}}]{Meng_Light_2021}%
  \BibitemOpen
  \bibfield  {author} {\bibinfo {author} {\bibfnamefont {Y.}~\bibnamefont {Meng}}, \bibinfo {author} {\bibfnamefont {Y.}~\bibnamefont {Chen}}, \bibinfo {author} {\bibfnamefont {L.}~\bibnamefont {Lu}}, \bibinfo {author} {\bibfnamefont {Y.}~\bibnamefont {Ding}}, \bibinfo {author} {\bibfnamefont {A.}~\bibnamefont {Cusano}}, \bibinfo {author} {\bibfnamefont {J.~A.}\ \bibnamefont {Fan}}, \bibinfo {author} {\bibfnamefont {Q.}~\bibnamefont {Hu}}, \bibinfo {author} {\bibfnamefont {K.}~\bibnamefont {Wang}}, \bibinfo {author} {\bibfnamefont {Z.}~\bibnamefont {Xie}}, \bibinfo {author} {\bibfnamefont {Z.}~\bibnamefont {Liu}}, \bibinfo {author} {\bibfnamefont {Y.}~\bibnamefont {Yang}}, \bibinfo {author} {\bibfnamefont {Q.}~\bibnamefont {Liu}}, \bibinfo {author} {\bibfnamefont {M.}~\bibnamefont {Gong}}, \bibinfo {author} {\bibfnamefont {Q.}~\bibnamefont {Xiao}}, \bibinfo {author} {\bibfnamefont {S.}~\bibnamefont {Sun}}, \bibinfo {author} {\bibfnamefont {M.}~\bibnamefont {Zhang}}, \bibinfo {author} {\bibfnamefont
  {X.}~\bibnamefont {Yuan}},\ and\ \bibinfo {author} {\bibfnamefont {X.}~\bibnamefont {Ni}},\ }\bibfield  {title} {\bibinfo {title} {Optical meta-waveguides for integrated photonics and beyond},\ }\href {https://doi.org/10.1038/s41377-021-00655-x} {\bibfield  {journal} {\bibinfo  {journal} {Light Sci. Appl.}\ }\textbf {\bibinfo {volume} {10}},\ \bibinfo {pages} {235} (\bibinfo {year} {2021})}\BibitemShut {NoStop}%
\bibitem [{\citenamefont {Snyder}\ and\ \citenamefont {Love}(1983)}]{snyder1983optical}%
  \BibitemOpen
  \bibfield  {author} {\bibinfo {author} {\bibfnamefont {A.~W.}\ \bibnamefont {Snyder}}\ and\ \bibinfo {author} {\bibfnamefont {J.~D.}\ \bibnamefont {Love}},\ }\href@noop {} {\emph {\bibinfo {title} {Optical Waveguide Theory}}}\ (\bibinfo  {publisher} {Chapman and Hall},\ \bibinfo {address} {London},\ \bibinfo {year} {1983})\BibitemShut {NoStop}%
\bibitem [{\citenamefont {Yermakov}\ \emph {et~al.}(2016)\citenamefont {Yermakov}, \citenamefont {Ovcharenko}, \citenamefont {Bogdanov}, \citenamefont {Iorsh}, \citenamefont {Bliokh},\ and\ \citenamefont {Kivshar}}]{yermakov2016spin}%
  \BibitemOpen
  \bibfield  {author} {\bibinfo {author} {\bibfnamefont {O.~Y.}\ \bibnamefont {Yermakov}}, \bibinfo {author} {\bibfnamefont {A.~I.}\ \bibnamefont {Ovcharenko}}, \bibinfo {author} {\bibfnamefont {A.~A.}\ \bibnamefont {Bogdanov}}, \bibinfo {author} {\bibfnamefont {I.~V.}\ \bibnamefont {Iorsh}}, \bibinfo {author} {\bibfnamefont {K.~Y.}\ \bibnamefont {Bliokh}},\ and\ \bibinfo {author} {\bibfnamefont {Y.~S.}\ \bibnamefont {Kivshar}},\ }\bibfield  {title} {\bibinfo {title} {Spin control of light with hyperbolic metasurfaces},\ }\href {https://doi.org/10.1103/PhysRevB.94.075446} {\bibfield  {journal} {\bibinfo  {journal} {Phys. Rev. B}\ }\textbf {\bibinfo {volume} {94}},\ \bibinfo {pages} {075446} (\bibinfo {year} {2016})}\BibitemShut {NoStop}%
\bibitem [{\citenamefont {Evlyukhin}\ \emph {et~al.}(2012)\citenamefont {Evlyukhin}, \citenamefont {Novikov}, \citenamefont {Zywietz}, \citenamefont {Eriksen}, \citenamefont {Reinhardt}, \citenamefont {Bozhevolnyi},\ and\ \citenamefont {Chichkov}}]{evlyukhin2012demonstration}%
  \BibitemOpen
  \bibfield  {author} {\bibinfo {author} {\bibfnamefont {A.~B.}\ \bibnamefont {Evlyukhin}}, \bibinfo {author} {\bibfnamefont {S.~M.}\ \bibnamefont {Novikov}}, \bibinfo {author} {\bibfnamefont {U.}~\bibnamefont {Zywietz}}, \bibinfo {author} {\bibfnamefont {R.~L.}\ \bibnamefont {Eriksen}}, \bibinfo {author} {\bibfnamefont {C.}~\bibnamefont {Reinhardt}}, \bibinfo {author} {\bibfnamefont {S.~I.}\ \bibnamefont {Bozhevolnyi}},\ and\ \bibinfo {author} {\bibfnamefont {B.~N.}\ \bibnamefont {Chichkov}},\ }\bibfield  {title} {\bibinfo {title} {Demonstration of magnetic dipole resonances of dielectric nanospheres in the visible region},\ }\href {https://doi.org/10.1021/nl301594s} {\bibfield  {journal} {\bibinfo  {journal} {Nano Lett.}\ }\textbf {\bibinfo {volume} {12}},\ \bibinfo {pages} {3749} (\bibinfo {year} {2012})}\BibitemShut {NoStop}%
\bibitem [{\citenamefont {Kuznetsov}\ \emph {et~al.}(2012)\citenamefont {Kuznetsov}, \citenamefont {Miroshnichenko}, \citenamefont {Fu}, \citenamefont {Zhang},\ and\ \citenamefont {Luk’Yanchuk}}]{kuznetsov2012magnetic}%
  \BibitemOpen
  \bibfield  {author} {\bibinfo {author} {\bibfnamefont {A.~I.}\ \bibnamefont {Kuznetsov}}, \bibinfo {author} {\bibfnamefont {A.~E.}\ \bibnamefont {Miroshnichenko}}, \bibinfo {author} {\bibfnamefont {Y.~H.}\ \bibnamefont {Fu}}, \bibinfo {author} {\bibfnamefont {J.}~\bibnamefont {Zhang}},\ and\ \bibinfo {author} {\bibfnamefont {B.}~\bibnamefont {Luk’Yanchuk}},\ }\bibfield  {title} {\bibinfo {title} {Magnetic light},\ }\href {https://doi.org/10.1038/srep00492} {\bibfield  {journal} {\bibinfo  {journal} {Sci. Rep.}\ }\textbf {\bibinfo {volume} {2}},\ \bibinfo {pages} {492} (\bibinfo {year} {2012})}\BibitemShut {NoStop}%
\bibitem [{\citenamefont {Kruk}\ and\ \citenamefont {Kivshar}(2017)}]{kruk2017functional}%
  \BibitemOpen
  \bibfield  {author} {\bibinfo {author} {\bibfnamefont {S.}~\bibnamefont {Kruk}}\ and\ \bibinfo {author} {\bibfnamefont {Y.}~\bibnamefont {Kivshar}},\ }\bibfield  {title} {\bibinfo {title} {Functional meta-optics and nanophotonics governed by {Mie} resonances},\ }\href {https://doi.org/10.1021/acsphotonics.7b01038} {\bibfield  {journal} {\bibinfo  {journal} {ACS Photonics}\ }\textbf {\bibinfo {volume} {4}},\ \bibinfo {pages} {2638} (\bibinfo {year} {2017})}\BibitemShut {NoStop}%
\bibitem [{\citenamefont {Yermakov}\ \emph {et~al.}(2021)\citenamefont {Yermakov}, \citenamefont {Lenets}, \citenamefont {Sayanskiy}, \citenamefont {Baena}, \citenamefont {Martini}, \citenamefont {Glybovski},\ and\ \citenamefont {Maci}}]{Yermakov_PhysRevX.11.031038}%
  \BibitemOpen
  \bibfield  {author} {\bibinfo {author} {\bibfnamefont {O.}~\bibnamefont {Yermakov}}, \bibinfo {author} {\bibfnamefont {V.}~\bibnamefont {Lenets}}, \bibinfo {author} {\bibfnamefont {A.}~\bibnamefont {Sayanskiy}}, \bibinfo {author} {\bibfnamefont {J.}~\bibnamefont {Baena}}, \bibinfo {author} {\bibfnamefont {E.}~\bibnamefont {Martini}}, \bibinfo {author} {\bibfnamefont {S.}~\bibnamefont {Glybovski}},\ and\ \bibinfo {author} {\bibfnamefont {S.}~\bibnamefont {Maci}},\ }\bibfield  {title} {\bibinfo {title} {Surface waves on self-complementary metasurfaces: All-frequency hyperbolicity, extreme canalization, and te-tm polarization degeneracy},\ }\href {https://doi.org/10.1103/PhysRevX.11.031038} {\bibfield  {journal} {\bibinfo  {journal} {Phys. Rev. X}\ }\textbf {\bibinfo {volume} {11}},\ \bibinfo {pages} {031038} (\bibinfo {year} {2021})}\BibitemShut {NoStop}%
\bibitem [{\citenamefont {Asadulina}\ \emph {et~al.}(2024)\citenamefont {Asadulina}, \citenamefont {Bogdanov},\ and\ \citenamefont {Yermakov}}]{Asadulina_lpor_2023}%
  \BibitemOpen
  \bibfield  {author} {\bibinfo {author} {\bibfnamefont {S.}~\bibnamefont {Asadulina}}, \bibinfo {author} {\bibfnamefont {A.}~\bibnamefont {Bogdanov}},\ and\ \bibinfo {author} {\bibfnamefont {O.}~\bibnamefont {Yermakov}},\ }\bibfield  {title} {\bibinfo {title} {All-dielectric meta-waveguides for flexible polarization control of guided light},\ }\href {https://doi.org/https://doi.org/10.1002/lpor.202300544} {\bibfield  {journal} {\bibinfo  {journal} {Laser Photon. Rev.}\ }\textbf {\bibinfo {volume} {n/a}},\ \bibinfo {pages} {2300544} (\bibinfo {year} {2024})}\BibitemShut {NoStop}%
\bibitem [{\citenamefont {Baranov}\ \emph {et~al.}(2017)\citenamefont {Baranov}, \citenamefont {Zuev}, \citenamefont {Lepeshov}, \citenamefont {Kotov}, \citenamefont {Krasnok}, \citenamefont {Evlyukhin},\ and\ \citenamefont {Chichkov}}]{Baranov_Optica_2017}%
  \BibitemOpen
  \bibfield  {author} {\bibinfo {author} {\bibfnamefont {D.~G.}\ \bibnamefont {Baranov}}, \bibinfo {author} {\bibfnamefont {D.~A.}\ \bibnamefont {Zuev}}, \bibinfo {author} {\bibfnamefont {S.~I.}\ \bibnamefont {Lepeshov}}, \bibinfo {author} {\bibfnamefont {O.~V.}\ \bibnamefont {Kotov}}, \bibinfo {author} {\bibfnamefont {A.~E.}\ \bibnamefont {Krasnok}}, \bibinfo {author} {\bibfnamefont {A.~B.}\ \bibnamefont {Evlyukhin}},\ and\ \bibinfo {author} {\bibfnamefont {B.~N.}\ \bibnamefont {Chichkov}},\ }\bibfield  {title} {\bibinfo {title} {All-dielectric nanophotonics: the quest for better materials and fabrication techniques},\ }\href {https://doi.org/10.1364/OPTICA.4.000814} {\bibfield  {journal} {\bibinfo  {journal} {Optica}\ }\textbf {\bibinfo {volume} {4}},\ \bibinfo {pages} {814} (\bibinfo {year} {2017})}\BibitemShut {NoStop}%
\bibitem [{\citenamefont {Su}\ \emph {et~al.}(2020)\citenamefont {Su}, \citenamefont {Zhang}, \citenamefont {Qiu}, \citenamefont {Guo},\ and\ \citenamefont {Sun}}]{Su_admt_2020}%
  \BibitemOpen
  \bibfield  {author} {\bibinfo {author} {\bibfnamefont {Y.}~\bibnamefont {Su}}, \bibinfo {author} {\bibfnamefont {Y.}~\bibnamefont {Zhang}}, \bibinfo {author} {\bibfnamefont {C.}~\bibnamefont {Qiu}}, \bibinfo {author} {\bibfnamefont {X.}~\bibnamefont {Guo}},\ and\ \bibinfo {author} {\bibfnamefont {L.}~\bibnamefont {Sun}},\ }\bibfield  {title} {\bibinfo {title} {Silicon photonic platform for passive waveguide devices: materials, fabrication, and applications},\ }\href {https://doi.org/https://doi.org/10.1002/admt.201901153} {\bibfield  {journal} {\bibinfo  {journal} {Adv. Mater. Technol.}\ }\textbf {\bibinfo {volume} {5}},\ \bibinfo {pages} {1901153} (\bibinfo {year} {2020})}\BibitemShut {NoStop}%
\bibitem [{\citenamefont {Aguilar}\ \emph {et~al.}(2019)\citenamefont {Aguilar}, \citenamefont {de~Castro}, \citenamefont {Godoy},\ and\ \citenamefont {Dias}}]{aguilar2019optoelectronic}%
  \BibitemOpen
  \bibfield  {author} {\bibinfo {author} {\bibfnamefont {O.}~\bibnamefont {Aguilar}}, \bibinfo {author} {\bibfnamefont {S.}~\bibnamefont {de~Castro}}, \bibinfo {author} {\bibfnamefont {M.~P.~F.}\ \bibnamefont {Godoy}},\ and\ \bibinfo {author} {\bibfnamefont {M.~R.~S.}\ \bibnamefont {Dias}},\ }\bibfield  {title} {\bibinfo {title} {Optoelectronic characterization of {Zn$_{1-x}$Cd${_x}$O} thin films as an alternative to photonic crystals in organic solar cells},\ }\href {https://doi.org/10.1364/OME.9.003638} {\bibfield  {journal} {\bibinfo  {journal} {Opt. Mater. Express}\ }\textbf {\bibinfo {volume} {9}},\ \bibinfo {pages} {3638} (\bibinfo {year} {2019})}\BibitemShut {NoStop}%
\bibitem [{\citenamefont {Beliaev}\ \emph {et~al.}(2022)\citenamefont {Beliaev}, \citenamefont {Shkondin}, \citenamefont {Lavrinenko},\ and\ \citenamefont {Takayama}}]{beliaev2022optical}%
  \BibitemOpen
  \bibfield  {author} {\bibinfo {author} {\bibfnamefont {L.~Y.}\ \bibnamefont {Beliaev}}, \bibinfo {author} {\bibfnamefont {E.}~\bibnamefont {Shkondin}}, \bibinfo {author} {\bibfnamefont {A.~V.}\ \bibnamefont {Lavrinenko}},\ and\ \bibinfo {author} {\bibfnamefont {O.}~\bibnamefont {Takayama}},\ }\bibfield  {title} {\bibinfo {title} {Optical, structural and composition properties of silicon nitride films deposited by reactive radio-frequency sputtering, low pressure and plasma-enhanced chemical vapor deposition},\ }\href {https://doi.org/10.1016/j.tsf.2022.139568} {\bibfield  {journal} {\bibinfo  {journal} {Thin Solid Films}\ }\textbf {\bibinfo {volume} {763}},\ \bibinfo {pages} {139568} (\bibinfo {year} {2022})}\BibitemShut {NoStop}%
\bibitem [{\citenamefont {Sarkar}\ \emph {et~al.}(2019)\citenamefont {Sarkar}, \citenamefont {Gupta}, \citenamefont {Kumar}, \citenamefont {Schubert}, \citenamefont {Probst}, \citenamefont {Joseph},\ and\ \citenamefont {K{\"o}nig}}]{sarkar2019hybridized}%
  \BibitemOpen
  \bibfield  {author} {\bibinfo {author} {\bibfnamefont {S.}~\bibnamefont {Sarkar}}, \bibinfo {author} {\bibfnamefont {V.}~\bibnamefont {Gupta}}, \bibinfo {author} {\bibfnamefont {M.}~\bibnamefont {Kumar}}, \bibinfo {author} {\bibfnamefont {J.}~\bibnamefont {Schubert}}, \bibinfo {author} {\bibfnamefont {P.~T.}\ \bibnamefont {Probst}}, \bibinfo {author} {\bibfnamefont {J.}~\bibnamefont {Joseph}},\ and\ \bibinfo {author} {\bibfnamefont {T.~A.~F.}\ \bibnamefont {K{\"o}nig}},\ }\bibfield  {title} {\bibinfo {title} {Hybridized guided-mode resonances via colloidal plasmonic self-assembled grating},\ }\href {https://doi.org/10.1021/acsami.8b20535} {\bibfield  {journal} {\bibinfo  {journal} {ACS Appl. Mater. Interfaces}\ }\textbf {\bibinfo {volume} {11}},\ \bibinfo {pages} {13752} (\bibinfo {year} {2019})}\BibitemShut {NoStop}%
\bibitem [{\citenamefont {Schinke}\ \emph {et~al.}(2015)\citenamefont {Schinke}, \citenamefont {Christian~Peest}, \citenamefont {Schmidt}, \citenamefont {Brendel}, \citenamefont {Bothe}, \citenamefont {Vogt}, \citenamefont {Kr{\"o}ger}, \citenamefont {Winter}, \citenamefont {Schirmacher}, \citenamefont {Lim}, \citenamefont {Nguyen},\ and\ \citenamefont {MacDonald}}]{schinke2015uncertainty}%
  \BibitemOpen
  \bibfield  {author} {\bibinfo {author} {\bibfnamefont {C.}~\bibnamefont {Schinke}}, \bibinfo {author} {\bibfnamefont {P.}~\bibnamefont {Christian~Peest}}, \bibinfo {author} {\bibfnamefont {J.}~\bibnamefont {Schmidt}}, \bibinfo {author} {\bibfnamefont {R.}~\bibnamefont {Brendel}}, \bibinfo {author} {\bibfnamefont {K.}~\bibnamefont {Bothe}}, \bibinfo {author} {\bibfnamefont {M.~R.}\ \bibnamefont {Vogt}}, \bibinfo {author} {\bibfnamefont {I.}~\bibnamefont {Kr{\"o}ger}}, \bibinfo {author} {\bibfnamefont {S.}~\bibnamefont {Winter}}, \bibinfo {author} {\bibfnamefont {A.}~\bibnamefont {Schirmacher}}, \bibinfo {author} {\bibfnamefont {S.}~\bibnamefont {Lim}}, \bibinfo {author} {\bibfnamefont {H.~T.}\ \bibnamefont {Nguyen}},\ and\ \bibinfo {author} {\bibfnamefont {D.}~\bibnamefont {MacDonald}},\ }\bibfield  {title} {\bibinfo {title} {Uncertainty analysis for the coefficient of band-to-band absorption of crystalline silicon},\ }\bibfield  {journal} {\bibinfo  {journal} {AIP Adv.}\ }\textbf {\bibinfo {volume} {5}},\
  \href {https://doi.org/10.1063/1.4923379} {10.1063/1.4923379} (\bibinfo {year} {2015})\BibitemShut {NoStop}%
\bibitem [{\citenamefont {Shkondin}\ \emph {et~al.}(2017)\citenamefont {Shkondin}, \citenamefont {Takayama}, \citenamefont {Panah}, \citenamefont {Liu}, \citenamefont {Larsen}, \citenamefont {Mar}, \citenamefont {Jensen},\ and\ \citenamefont {Lavrinenko}}]{shkondin2017large}%
  \BibitemOpen
  \bibfield  {author} {\bibinfo {author} {\bibfnamefont {E.}~\bibnamefont {Shkondin}}, \bibinfo {author} {\bibfnamefont {O.}~\bibnamefont {Takayama}}, \bibinfo {author} {\bibfnamefont {M.~E.~A.}\ \bibnamefont {Panah}}, \bibinfo {author} {\bibfnamefont {P.}~\bibnamefont {Liu}}, \bibinfo {author} {\bibfnamefont {P.~V.}\ \bibnamefont {Larsen}}, \bibinfo {author} {\bibfnamefont {M.~D.}\ \bibnamefont {Mar}}, \bibinfo {author} {\bibfnamefont {F.}~\bibnamefont {Jensen}},\ and\ \bibinfo {author} {\bibfnamefont {A.~V.}\ \bibnamefont {Lavrinenko}},\ }\bibfield  {title} {\bibinfo {title} {Large-scale high aspect ratio {Al}-doped {ZnO} nanopillars arrays as anisotropic metamaterials},\ }\href {https://doi.org/10.1364/OME.7.001606} {\bibfield  {journal} {\bibinfo  {journal} {Opt. Mater. Express}\ }\textbf {\bibinfo {volume} {7}},\ \bibinfo {pages} {1606} (\bibinfo {year} {2017})}\BibitemShut {NoStop}%
\bibitem [{\citenamefont {Johnson}\ and\ \citenamefont {Joannopoulos}(2001)}]{johnson2001block}%
  \BibitemOpen
  \bibfield  {author} {\bibinfo {author} {\bibfnamefont {S.~G.}\ \bibnamefont {Johnson}}\ and\ \bibinfo {author} {\bibfnamefont {J.~D.}\ \bibnamefont {Joannopoulos}},\ }\bibfield  {title} {\bibinfo {title} {Block-iterative frequency-domain methods for {Maxwell's} equations in a planewave basis},\ }\href {https://doi.org/10.1364/OE.8.000173} {\bibfield  {journal} {\bibinfo  {journal} {Opt. Express}\ }\textbf {\bibinfo {volume} {8}},\ \bibinfo {pages} {173} (\bibinfo {year} {2001})}\BibitemShut {NoStop}%
\bibitem [{\citenamefont {Rodr{\'\i}guez-Fortu{\~n}o}\ \emph {et~al.}(2013)\citenamefont {Rodr{\'\i}guez-Fortu{\~n}o}, \citenamefont {Marino}, \citenamefont {Ginzburg}, \citenamefont {O’Connor}, \citenamefont {Mart{\'\i}nez}, \citenamefont {Wurtz},\ and\ \citenamefont {Zayats}}]{rodriguez2013near}%
  \BibitemOpen
  \bibfield  {author} {\bibinfo {author} {\bibfnamefont {F.~J.}\ \bibnamefont {Rodr{\'\i}guez-Fortu{\~n}o}}, \bibinfo {author} {\bibfnamefont {G.}~\bibnamefont {Marino}}, \bibinfo {author} {\bibfnamefont {P.}~\bibnamefont {Ginzburg}}, \bibinfo {author} {\bibfnamefont {D.}~\bibnamefont {O’Connor}}, \bibinfo {author} {\bibfnamefont {A.}~\bibnamefont {Mart{\'\i}nez}}, \bibinfo {author} {\bibfnamefont {G.~A.}\ \bibnamefont {Wurtz}},\ and\ \bibinfo {author} {\bibfnamefont {A.~V.}\ \bibnamefont {Zayats}},\ }\bibfield  {title} {\bibinfo {title} {Near-field interference for the unidirectional excitation of electromagnetic guided modes},\ }\href@noop {} {\bibfield  {journal} {\bibinfo  {journal} {Science}\ }\textbf {\bibinfo {volume} {340}},\ \bibinfo {pages} {328} (\bibinfo {year} {2013})}\BibitemShut {NoStop}%
\bibitem [{\citenamefont {Li}\ \emph {et~al.}(2015)\citenamefont {Li}, \citenamefont {Baranov}, \citenamefont {Krasnok},\ and\ \citenamefont {Belov}}]{li2015all}%
  \BibitemOpen
  \bibfield  {author} {\bibinfo {author} {\bibfnamefont {S.~V.}\ \bibnamefont {Li}}, \bibinfo {author} {\bibfnamefont {D.~G.}\ \bibnamefont {Baranov}}, \bibinfo {author} {\bibfnamefont {A.~E.}\ \bibnamefont {Krasnok}},\ and\ \bibinfo {author} {\bibfnamefont {P.~A.}\ \bibnamefont {Belov}},\ }\bibfield  {title} {\bibinfo {title} {All-dielectric nanoantennas for unidirectional excitation of electromagnetic guided modes},\ }\href@noop {} {\bibfield  {journal} {\bibinfo  {journal} {Appl. Phys. Lett.}\ }\textbf {\bibinfo {volume} {107}} (\bibinfo {year} {2015})}\BibitemShut {NoStop}%
\bibitem [{\citenamefont {Decker}\ \emph {et~al.}(2015)\citenamefont {Decker}, \citenamefont {Staude}, \citenamefont {Falkner}, \citenamefont {Dominguez}, \citenamefont {Neshev}, \citenamefont {Brener}, \citenamefont {Pertsch},\ and\ \citenamefont {Kivshar}}]{decker2015high}%
  \BibitemOpen
  \bibfield  {author} {\bibinfo {author} {\bibfnamefont {M.}~\bibnamefont {Decker}}, \bibinfo {author} {\bibfnamefont {I.}~\bibnamefont {Staude}}, \bibinfo {author} {\bibfnamefont {M.}~\bibnamefont {Falkner}}, \bibinfo {author} {\bibfnamefont {J.}~\bibnamefont {Dominguez}}, \bibinfo {author} {\bibfnamefont {D.~N.}\ \bibnamefont {Neshev}}, \bibinfo {author} {\bibfnamefont {I.}~\bibnamefont {Brener}}, \bibinfo {author} {\bibfnamefont {T.}~\bibnamefont {Pertsch}},\ and\ \bibinfo {author} {\bibfnamefont {Y.~S.}\ \bibnamefont {Kivshar}},\ }\bibfield  {title} {\bibinfo {title} {High-efficiency dielectric {Huygens’} surfaces},\ }\href@noop {} {\bibfield  {journal} {\bibinfo  {journal} {Adv. Opt. Mater.}\ }\textbf {\bibinfo {volume} {3}},\ \bibinfo {pages} {813} (\bibinfo {year} {2015})}\BibitemShut {NoStop}%
\bibitem [{\citenamefont {Chong}\ \emph {et~al.}(2016)\citenamefont {Chong}, \citenamefont {Wang}, \citenamefont {Staude}, \citenamefont {James}, \citenamefont {Dominguez}, \citenamefont {Liu}, \citenamefont {Subramania}, \citenamefont {Decker}, \citenamefont {Neshev}, \citenamefont {Brener} \emph {et~al.}}]{chong2016efficient}%
  \BibitemOpen
  \bibfield  {author} {\bibinfo {author} {\bibfnamefont {K.~E.}\ \bibnamefont {Chong}}, \bibinfo {author} {\bibfnamefont {L.}~\bibnamefont {Wang}}, \bibinfo {author} {\bibfnamefont {I.}~\bibnamefont {Staude}}, \bibinfo {author} {\bibfnamefont {A.~R.}\ \bibnamefont {James}}, \bibinfo {author} {\bibfnamefont {J.}~\bibnamefont {Dominguez}}, \bibinfo {author} {\bibfnamefont {S.}~\bibnamefont {Liu}}, \bibinfo {author} {\bibfnamefont {G.~S.}\ \bibnamefont {Subramania}}, \bibinfo {author} {\bibfnamefont {M.}~\bibnamefont {Decker}}, \bibinfo {author} {\bibfnamefont {D.~N.}\ \bibnamefont {Neshev}}, \bibinfo {author} {\bibfnamefont {I.}~\bibnamefont {Brener}}, \emph {et~al.},\ }\bibfield  {title} {\bibinfo {title} {Efficient polarization-insensitive complex wavefront control using huygens’ metasurfaces based on dielectric resonant meta-atoms},\ }\href@noop {} {\bibfield  {journal} {\bibinfo  {journal} {ACS Photonics}\ }\textbf {\bibinfo {volume} {3}},\ \bibinfo {pages} {514} (\bibinfo {year} {2016})}\BibitemShut
  {NoStop}%
\bibitem [{\citenamefont {Prokhorov}\ \emph {et~al.}(2022)\citenamefont {Prokhorov}, \citenamefont {Terekhov}, \citenamefont {Gubin}, \citenamefont {Shesterikov}, \citenamefont {Ni}, \citenamefont {Tuz},\ and\ \citenamefont {Evlyukhin}}]{Evlyukhin_acsphotonics_2022}%
  \BibitemOpen
  \bibfield  {author} {\bibinfo {author} {\bibfnamefont {A.~V.}\ \bibnamefont {Prokhorov}}, \bibinfo {author} {\bibfnamefont {P.~D.}\ \bibnamefont {Terekhov}}, \bibinfo {author} {\bibfnamefont {M.~Y.}\ \bibnamefont {Gubin}}, \bibinfo {author} {\bibfnamefont {A.~V.}\ \bibnamefont {Shesterikov}}, \bibinfo {author} {\bibfnamefont {X.}~\bibnamefont {Ni}}, \bibinfo {author} {\bibfnamefont {V.~R.}\ \bibnamefont {Tuz}},\ and\ \bibinfo {author} {\bibfnamefont {A.~B.}\ \bibnamefont {Evlyukhin}},\ }\bibfield  {title} {\bibinfo {title} {Resonant light trapping via lattice-induced multipole coupling in symmetrical metasurfaces},\ }\href {https://doi.org/10.1021/acsphotonics.2c01066} {\bibfield  {journal} {\bibinfo  {journal} {ACS Photonics}\ }\textbf {\bibinfo {volume} {9}},\ \bibinfo {pages} {3869} (\bibinfo {year} {2022})}\BibitemShut {NoStop}%
\bibitem [{\citenamefont {Allayarov}\ \emph {et~al.}(2024)\citenamefont {Allayarov}, \citenamefont {Evlyukhin},\ and\ \citenamefont {Lesina}}]{Allayarov_OptExpress_2024}%
  \BibitemOpen
  \bibfield  {author} {\bibinfo {author} {\bibfnamefont {I.}~\bibnamefont {Allayarov}}, \bibinfo {author} {\bibfnamefont {A.~B.}\ \bibnamefont {Evlyukhin}},\ and\ \bibinfo {author} {\bibfnamefont {A.~C.}\ \bibnamefont {Lesina}},\ }\bibfield  {title} {\bibinfo {title} {Multiresonant all-dielectric metasurfaces based on high-order multipole coupling in the visible},\ }\href {https://doi.org/10.1364/OE.511172} {\bibfield  {journal} {\bibinfo  {journal} {Opt. Express}\ }\textbf {\bibinfo {volume} {32}},\ \bibinfo {pages} {5641} (\bibinfo {year} {2024})}\BibitemShut {NoStop}%
\bibitem [{\citenamefont {Mongia}\ and\ \citenamefont {Bhartia}(1994)}]{Mongia_1994}%
  \BibitemOpen
  \bibfield  {author} {\bibinfo {author} {\bibfnamefont {R.~K.}\ \bibnamefont {Mongia}}\ and\ \bibinfo {author} {\bibfnamefont {P.}~\bibnamefont {Bhartia}},\ }\bibfield  {title} {\bibinfo {title} {Dielectric resonator antennas—a review and general design relations for resonant frequency and bandwidth},\ }\href {https://doi.org/10.1002/mmce.4570040304} {\bibfield  {journal} {\bibinfo  {journal} {Int. J. RF Microw. Comput. Aided Eng.}\ }\textbf {\bibinfo {volume} {4}},\ \bibinfo {pages} {230} (\bibinfo {year} {1994})}\BibitemShut {NoStop}%
\bibitem [{\citenamefont {Babicheva}\ and\ \citenamefont {Evlyukhin}(2017)}]{babicheva2017resonant}%
  \BibitemOpen
  \bibfield  {author} {\bibinfo {author} {\bibfnamefont {V.~E.}\ \bibnamefont {Babicheva}}\ and\ \bibinfo {author} {\bibfnamefont {A.~B.}\ \bibnamefont {Evlyukhin}},\ }\bibfield  {title} {\bibinfo {title} {Resonant lattice {Kerker} effect in metasurfaces with electric and magnetic optical responses},\ }\href@noop {} {\bibfield  {journal} {\bibinfo  {journal} {Las. Photon. Rev.}\ }\textbf {\bibinfo {volume} {11}},\ \bibinfo {pages} {1700132} (\bibinfo {year} {2017})}\BibitemShut {NoStop}%
\bibitem [{\citenamefont {Shamkhi}\ \emph {et~al.}(2019{\natexlab{a}})\citenamefont {Shamkhi}, \citenamefont {Baryshnikova}, \citenamefont {Sayanskiy}, \citenamefont {Kapitanova}, \citenamefont {Terekhov}, \citenamefont {Belov}, \citenamefont {Karabchevsky}, \citenamefont {Evlyukhin}, \citenamefont {Kivshar},\ and\ \citenamefont {Shalin}}]{Shamkhi_PhysRevLett_2019}%
  \BibitemOpen
  \bibfield  {author} {\bibinfo {author} {\bibfnamefont {H.~K.}\ \bibnamefont {Shamkhi}}, \bibinfo {author} {\bibfnamefont {K.~V.}\ \bibnamefont {Baryshnikova}}, \bibinfo {author} {\bibfnamefont {A.}~\bibnamefont {Sayanskiy}}, \bibinfo {author} {\bibfnamefont {P.}~\bibnamefont {Kapitanova}}, \bibinfo {author} {\bibfnamefont {P.~D.}\ \bibnamefont {Terekhov}}, \bibinfo {author} {\bibfnamefont {P.}~\bibnamefont {Belov}}, \bibinfo {author} {\bibfnamefont {A.}~\bibnamefont {Karabchevsky}}, \bibinfo {author} {\bibfnamefont {A.~B.}\ \bibnamefont {Evlyukhin}}, \bibinfo {author} {\bibfnamefont {Y.}~\bibnamefont {Kivshar}},\ and\ \bibinfo {author} {\bibfnamefont {A.~S.}\ \bibnamefont {Shalin}},\ }\bibfield  {title} {\bibinfo {title} {Transverse scattering and generalized {Kerker} effects in all-dielectric {Mie}-resonant metaoptics},\ }\href {https://doi.org/10.1103/PhysRevLett.122.193905} {\bibfield  {journal} {\bibinfo  {journal} {Phys. Rev. Lett.}\ }\textbf {\bibinfo {volume} {122}},\ \bibinfo {pages} {193905}
  (\bibinfo {year} {2019}{\natexlab{a}})}\BibitemShut {NoStop}%
\bibitem [{\citenamefont {Shamkhi}\ \emph {et~al.}(2019{\natexlab{b}})\citenamefont {Shamkhi}, \citenamefont {Sayanskiy}, \citenamefont {Valero}, \citenamefont {Kupriianov}, \citenamefont {Kapitanova}, \citenamefont {Kivshar}, \citenamefont {Shalin},\ and\ \citenamefont {Tuz}}]{Shamkhi_PhysRevMaterials_2019}%
  \BibitemOpen
  \bibfield  {author} {\bibinfo {author} {\bibfnamefont {H.~K.}\ \bibnamefont {Shamkhi}}, \bibinfo {author} {\bibfnamefont {A.}~\bibnamefont {Sayanskiy}}, \bibinfo {author} {\bibfnamefont {A.~C.}\ \bibnamefont {Valero}}, \bibinfo {author} {\bibfnamefont {A.~S.}\ \bibnamefont {Kupriianov}}, \bibinfo {author} {\bibfnamefont {P.}~\bibnamefont {Kapitanova}}, \bibinfo {author} {\bibfnamefont {Y.~S.}\ \bibnamefont {Kivshar}}, \bibinfo {author} {\bibfnamefont {A.~S.}\ \bibnamefont {Shalin}},\ and\ \bibinfo {author} {\bibfnamefont {V.~R.}\ \bibnamefont {Tuz}},\ }\bibfield  {title} {\bibinfo {title} {Transparency and perfect absorption of all-dielectric resonant metasurfaces governed by the transverse {Kerker} effect},\ }\href {https://doi.org/10.1103/PhysRevMaterials.3.085201} {\bibfield  {journal} {\bibinfo  {journal} {Phys. Rev. Mater.}\ }\textbf {\bibinfo {volume} {3}},\ \bibinfo {pages} {085201} (\bibinfo {year} {2019}{\natexlab{b}})}\BibitemShut {NoStop}%
\bibitem [{\citenamefont {Colburn}\ \emph {et~al.}(2018)\citenamefont {Colburn}, \citenamefont {Zhan}, \citenamefont {Bayati}, \citenamefont {Whitehead}, \citenamefont {Ryou}, \citenamefont {Huang},\ and\ \citenamefont {Majumdar}}]{colburn2018broadband}%
  \BibitemOpen
  \bibfield  {author} {\bibinfo {author} {\bibfnamefont {S.}~\bibnamefont {Colburn}}, \bibinfo {author} {\bibfnamefont {A.}~\bibnamefont {Zhan}}, \bibinfo {author} {\bibfnamefont {E.}~\bibnamefont {Bayati}}, \bibinfo {author} {\bibfnamefont {J.}~\bibnamefont {Whitehead}}, \bibinfo {author} {\bibfnamefont {A.}~\bibnamefont {Ryou}}, \bibinfo {author} {\bibfnamefont {L.}~\bibnamefont {Huang}},\ and\ \bibinfo {author} {\bibfnamefont {A.}~\bibnamefont {Majumdar}},\ }\bibfield  {title} {\bibinfo {title} {Broadband transparent and {CMOS}-compatible flat optics with silicon nitride metasurfaces},\ }\href {https://doi.org/10.1364/OME.8.002330} {\bibfield  {journal} {\bibinfo  {journal} {Opt. Mater. Express}\ }\textbf {\bibinfo {volume} {8}},\ \bibinfo {pages} {2330} (\bibinfo {year} {2018})}\BibitemShut {NoStop}%
\bibitem [{\citenamefont {Yang}\ \emph {et~al.}(2020)\citenamefont {Yang}, \citenamefont {Babicheva}, \citenamefont {Yu}, \citenamefont {Lu}, \citenamefont {Lin},\ and\ \citenamefont {Chen}}]{yang2020structural}%
  \BibitemOpen
  \bibfield  {author} {\bibinfo {author} {\bibfnamefont {J.-H.}\ \bibnamefont {Yang}}, \bibinfo {author} {\bibfnamefont {V.~E.}\ \bibnamefont {Babicheva}}, \bibinfo {author} {\bibfnamefont {M.-W.}\ \bibnamefont {Yu}}, \bibinfo {author} {\bibfnamefont {T.-C.}\ \bibnamefont {Lu}}, \bibinfo {author} {\bibfnamefont {T.-R.}\ \bibnamefont {Lin}},\ and\ \bibinfo {author} {\bibfnamefont {K.-P.}\ \bibnamefont {Chen}},\ }\bibfield  {title} {\bibinfo {title} {Structural colors enabled by lattice resonance on silicon nitride metasurfaces},\ }\href {https://doi.org/10.1021/acsnano.0c00185} {\bibfield  {journal} {\bibinfo  {journal} {ACS Nano}\ }\textbf {\bibinfo {volume} {14}},\ \bibinfo {pages} {5678} (\bibinfo {year} {2020})}\BibitemShut {NoStop}%
\bibitem [{\citenamefont {Sun}\ \emph {et~al.}(2021)\citenamefont {Sun}, \citenamefont {Yu}, \citenamefont {Deng}, \citenamefont {Ban}, \citenamefont {Liu},\ and\ \citenamefont {Qiu}}]{sun2021electro}%
  \BibitemOpen
  \bibfield  {author} {\bibinfo {author} {\bibfnamefont {X.}~\bibnamefont {Sun}}, \bibinfo {author} {\bibfnamefont {H.}~\bibnamefont {Yu}}, \bibinfo {author} {\bibfnamefont {N.}~\bibnamefont {Deng}}, \bibinfo {author} {\bibfnamefont {D.}~\bibnamefont {Ban}}, \bibinfo {author} {\bibfnamefont {G.}~\bibnamefont {Liu}},\ and\ \bibinfo {author} {\bibfnamefont {F.}~\bibnamefont {Qiu}},\ }\bibfield  {title} {\bibinfo {title} {Electro-optic polymer and silicon nitride hybrid spatial light modulators based on a metasurface},\ }\href {https://doi.org/10.1364/OE.434480} {\bibfield  {journal} {\bibinfo  {journal} {Opt. Express}\ }\textbf {\bibinfo {volume} {29}},\ \bibinfo {pages} {25543} (\bibinfo {year} {2021})}\BibitemShut {NoStop}%
\bibitem [{\citenamefont {Yermakov}\ \emph {et~al.}(2020)\citenamefont {Yermakov}, \citenamefont {Schneidewind}, \citenamefont {H{\"u}bner}, \citenamefont {Wieduwilt}, \citenamefont {Zeisberger}, \citenamefont {Bogdanov}, \citenamefont {Kivshar},\ and\ \citenamefont {Schmidt}}]{yermakov2020nanostructure}%
  \BibitemOpen
  \bibfield  {author} {\bibinfo {author} {\bibfnamefont {O.}~\bibnamefont {Yermakov}}, \bibinfo {author} {\bibfnamefont {H.}~\bibnamefont {Schneidewind}}, \bibinfo {author} {\bibfnamefont {U.}~\bibnamefont {H{\"u}bner}}, \bibinfo {author} {\bibfnamefont {T.}~\bibnamefont {Wieduwilt}}, \bibinfo {author} {\bibfnamefont {M.}~\bibnamefont {Zeisberger}}, \bibinfo {author} {\bibfnamefont {A.}~\bibnamefont {Bogdanov}}, \bibinfo {author} {\bibfnamefont {Y.}~\bibnamefont {Kivshar}},\ and\ \bibinfo {author} {\bibfnamefont {M.~A.}\ \bibnamefont {Schmidt}},\ }\bibfield  {title} {\bibinfo {title} {Nanostructure-empowered efficient coupling of light into optical fibers at extraordinarily large angles},\ }\href {https://doi.org/10.1021/acsphotonics.0c01078} {\bibfield  {journal} {\bibinfo  {journal} {ACS Photonics}\ }\textbf {\bibinfo {volume} {7}},\ \bibinfo {pages} {2834} (\bibinfo {year} {2020})}\BibitemShut {NoStop}%
\bibitem [{\citenamefont {Subramanian}\ \emph {et~al.}(2015)\citenamefont {Subramanian}, \citenamefont {Ryckeboer}, \citenamefont {Dhakal}, \citenamefont {Peyskens}, \citenamefont {Malik}, \citenamefont {Kuyken}, \citenamefont {Zhao}, \citenamefont {Pathak}, \citenamefont {Ruocco}, \citenamefont {De~Groote}, \citenamefont {Wuytens}, \citenamefont {Martens}, \citenamefont {Leo}, \citenamefont {Xie}, \citenamefont {Dave}, \citenamefont {Muneeb}, \citenamefont {Van~Dorpe}, \citenamefont {Van~Campenhout}, \citenamefont {Bogaerts}, \citenamefont {Bienstman}, \citenamefont {Le~Thomas}, \citenamefont {Van~Thourhout}, \citenamefont {Hens}, \citenamefont {Roelkens},\ and\ \citenamefont {Baets}}]{Subramanian_PhotonRes_2015}%
  \BibitemOpen
  \bibfield  {author} {\bibinfo {author} {\bibfnamefont {A.~Z.}\ \bibnamefont {Subramanian}}, \bibinfo {author} {\bibfnamefont {E.}~\bibnamefont {Ryckeboer}}, \bibinfo {author} {\bibfnamefont {A.}~\bibnamefont {Dhakal}}, \bibinfo {author} {\bibfnamefont {F.}~\bibnamefont {Peyskens}}, \bibinfo {author} {\bibfnamefont {A.}~\bibnamefont {Malik}}, \bibinfo {author} {\bibfnamefont {B.}~\bibnamefont {Kuyken}}, \bibinfo {author} {\bibfnamefont {H.}~\bibnamefont {Zhao}}, \bibinfo {author} {\bibfnamefont {S.}~\bibnamefont {Pathak}}, \bibinfo {author} {\bibfnamefont {A.}~\bibnamefont {Ruocco}}, \bibinfo {author} {\bibfnamefont {A.}~\bibnamefont {De~Groote}}, \bibinfo {author} {\bibfnamefont {P.}~\bibnamefont {Wuytens}}, \bibinfo {author} {\bibfnamefont {D.}~\bibnamefont {Martens}}, \bibinfo {author} {\bibfnamefont {F.}~\bibnamefont {Leo}}, \bibinfo {author} {\bibfnamefont {W.}~\bibnamefont {Xie}}, \bibinfo {author} {\bibfnamefont {U.~D.}\ \bibnamefont {Dave}}, \bibinfo {author} {\bibfnamefont {M.}~\bibnamefont
  {Muneeb}}, \bibinfo {author} {\bibfnamefont {P.}~\bibnamefont {Van~Dorpe}}, \bibinfo {author} {\bibfnamefont {J.}~\bibnamefont {Van~Campenhout}}, \bibinfo {author} {\bibfnamefont {W.}~\bibnamefont {Bogaerts}}, \bibinfo {author} {\bibfnamefont {P.}~\bibnamefont {Bienstman}}, \bibinfo {author} {\bibfnamefont {N.}~\bibnamefont {Le~Thomas}}, \bibinfo {author} {\bibfnamefont {D.}~\bibnamefont {Van~Thourhout}}, \bibinfo {author} {\bibfnamefont {Z.}~\bibnamefont {Hens}}, \bibinfo {author} {\bibfnamefont {G.}~\bibnamefont {Roelkens}},\ and\ \bibinfo {author} {\bibfnamefont {R.}~\bibnamefont {Baets}},\ }\bibfield  {title} {\bibinfo {title} {Silicon and silicon nitride photonic circuits for spectroscopic sensing on-a-chip {[Invited]}},\ }\href {https://doi.org/10.1364/PRJ.3.000B47} {\bibfield  {journal} {\bibinfo  {journal} {Photon. Res.}\ }\textbf {\bibinfo {volume} {3}},\ \bibinfo {pages} {B47} (\bibinfo {year} {2015})}\BibitemShut {NoStop}%
\bibitem [{\citenamefont {Malitson}(1965)}]{malitson1965interspecimen}%
  \BibitemOpen
  \bibfield  {author} {\bibinfo {author} {\bibfnamefont {I.~H.}\ \bibnamefont {Malitson}},\ }\bibfield  {title} {\bibinfo {title} {Interspecimen comparison of the refractive index of fused silica},\ }\href {https://doi.org/10.1364/JOSA.55.001205} {\bibfield  {journal} {\bibinfo  {journal} {J. Opt. Soc. Am.}\ }\textbf {\bibinfo {volume} {55}},\ \bibinfo {pages} {1205} (\bibinfo {year} {1965})}\BibitemShut {NoStop}%
\bibitem [{\citenamefont {Filonov}\ \emph {et~al.}(2012)\citenamefont {Filonov}, \citenamefont {Krasnok}, \citenamefont {Slobozhanyuk}, \citenamefont {Kapitanova}, \citenamefont {Nenasheva}, \citenamefont {Kivshar},\ and\ \citenamefont {Belov}}]{Filonov_ApplPhysLett_2012}%
  \BibitemOpen
  \bibfield  {author} {\bibinfo {author} {\bibfnamefont {D.~S.}\ \bibnamefont {Filonov}}, \bibinfo {author} {\bibfnamefont {A.~E.}\ \bibnamefont {Krasnok}}, \bibinfo {author} {\bibfnamefont {A.~P.}\ \bibnamefont {Slobozhanyuk}}, \bibinfo {author} {\bibfnamefont {P.~V.}\ \bibnamefont {Kapitanova}}, \bibinfo {author} {\bibfnamefont {E.~A.}\ \bibnamefont {Nenasheva}}, \bibinfo {author} {\bibfnamefont {Y.~S.}\ \bibnamefont {Kivshar}},\ and\ \bibinfo {author} {\bibfnamefont {P.~A.}\ \bibnamefont {Belov}},\ }\bibfield  {title} {\bibinfo {title} {Experimental verification of the concept of all-dielectric nanoantennas},\ }\href {https://doi.org/10.1063/1.4719209} {\bibfield  {journal} {\bibinfo  {journal} {Appl. Phys. Lett.}\ }\textbf {\bibinfo {volume} {100}},\ \bibinfo {pages} {201113} (\bibinfo {year} {2012})}\BibitemShut {NoStop}%
\bibitem [{\citenamefont {Xu}\ \emph {et~al.}(2019)\citenamefont {Xu}, \citenamefont {Sayanskiy}, \citenamefont {Kupriianov}, \citenamefont {Tuz}, \citenamefont {Kapitanova}, \citenamefont {Sun}, \citenamefont {Han},\ and\ \citenamefont {Kivshar}}]{Xu_AdvOptMater_2019}%
  \BibitemOpen
  \bibfield  {author} {\bibinfo {author} {\bibfnamefont {S.}~\bibnamefont {Xu}}, \bibinfo {author} {\bibfnamefont {A.}~\bibnamefont {Sayanskiy}}, \bibinfo {author} {\bibfnamefont {A.~S.}\ \bibnamefont {Kupriianov}}, \bibinfo {author} {\bibfnamefont {V.~R.}\ \bibnamefont {Tuz}}, \bibinfo {author} {\bibfnamefont {P.}~\bibnamefont {Kapitanova}}, \bibinfo {author} {\bibfnamefont {H.-B.}\ \bibnamefont {Sun}}, \bibinfo {author} {\bibfnamefont {W.}~\bibnamefont {Han}},\ and\ \bibinfo {author} {\bibfnamefont {Y.~S.}\ \bibnamefont {Kivshar}},\ }\bibfield  {title} {\bibinfo {title} {Experimental observation of toroidal dipole modes in all-dielectric metasurfaces},\ }\href {https://doi.org/https://doi.org/10.1002/adom.201801166} {\bibfield  {journal} {\bibinfo  {journal} {Adv. Opt. Mater.}\ }\textbf {\bibinfo {volume} {7}},\ \bibinfo {pages} {1801166} (\bibinfo {year} {2019})}\BibitemShut {NoStop}%
\bibitem [{\citenamefont {Kupriianov}\ \emph {et~al.}(2023)\citenamefont {Kupriianov}, \citenamefont {Khardikov}, \citenamefont {Domina}, \citenamefont {Prosvirnin}, \citenamefont {Han},\ and\ \citenamefont {Tuz}}]{Kupriianov_JApplPhys_2023}%
  \BibitemOpen
  \bibfield  {author} {\bibinfo {author} {\bibfnamefont {A.~S.}\ \bibnamefont {Kupriianov}}, \bibinfo {author} {\bibfnamefont {V.~V.}\ \bibnamefont {Khardikov}}, \bibinfo {author} {\bibfnamefont {K.}~\bibnamefont {Domina}}, \bibinfo {author} {\bibfnamefont {S.~L.}\ \bibnamefont {Prosvirnin}}, \bibinfo {author} {\bibfnamefont {W.}~\bibnamefont {Han}},\ and\ \bibinfo {author} {\bibfnamefont {V.~R.}\ \bibnamefont {Tuz}},\ }\bibfield  {title} {\bibinfo {title} {Experimental observation of diffractive retroreflection from a dielectric metasurface},\ }\href {https://doi.org/10.1063/5.0145338} {\bibfield  {journal} {\bibinfo  {journal} {J. Appl. Phys.}\ }\textbf {\bibinfo {volume} {133}},\ \bibinfo {pages} {163101} (\bibinfo {year} {2023})}\BibitemShut {NoStop}%
\bibitem [{\citenamefont {Ubic}\ \emph {et~al.}(2017)\citenamefont {Ubic}, \citenamefont {Subodh},\ and\ \citenamefont {Sebastian}}]{materials_2017}%
  \BibitemOpen
  \bibfield  {author} {\bibinfo {author} {\bibfnamefont {R.}~\bibnamefont {Ubic}}, \bibinfo {author} {\bibfnamefont {G.}~\bibnamefont {Subodh}},\ and\ \bibinfo {author} {\bibfnamefont {M.~T.}\ \bibnamefont {Sebastian}},\ }\bibfield  {title} {\bibinfo {title} {High permittivity materials},\ }in\ \href@noop {} {\emph {\bibinfo {booktitle} {Microwave Materials and Applications}}},\ Vol.~\bibinfo {volume} {1},\ \bibinfo {editor} {edited by\ \bibinfo {editor} {\bibfnamefont {M.~T.}\ \bibnamefont {Sebastian}}, \bibinfo {editor} {\bibfnamefont {R.}~\bibnamefont {Ubic}},\ and\ \bibinfo {editor} {\bibfnamefont {H.}~\bibnamefont {Jantunen}}}\ (\bibinfo  {publisher} {John Wiley \& Sons},\ \bibinfo {address} {Hoboken, NJ},\ \bibinfo {year} {2017})\ Chap.~\bibinfo {chapter} {4}, pp.\ \bibinfo {pages} {149--202}\BibitemShut {NoStop}%
\bibitem [{\citenamefont {Wangling}()}]{wangling}%
  \BibitemOpen
  \bibfield  {author} {\bibinfo {author} {\bibnamefont {Wangling}},\ }\href@noop {} {\bibinfo {title} {Data sheet of microwave thermoplastic material}},\ \bibinfo {howpublished} {\url{http://www.wang-ling.com.cn/index-en.html}}\BibitemShut {NoStop}%
\bibitem [{\citenamefont {Dockrey}\ \emph {et~al.}(2016)\citenamefont {Dockrey}, \citenamefont {Horsley}, \citenamefont {Hooper}, \citenamefont {Sambles},\ and\ \citenamefont {Hibbins}}]{dockrey2016direct}%
  \BibitemOpen
  \bibfield  {author} {\bibinfo {author} {\bibfnamefont {J.~A.}\ \bibnamefont {Dockrey}}, \bibinfo {author} {\bibfnamefont {S.~A.~R.}\ \bibnamefont {Horsley}}, \bibinfo {author} {\bibfnamefont {I.~R.}\ \bibnamefont {Hooper}}, \bibinfo {author} {\bibfnamefont {J.~R.}\ \bibnamefont {Sambles}},\ and\ \bibinfo {author} {\bibfnamefont {A.~P.}\ \bibnamefont {Hibbins}},\ }\bibfield  {title} {\bibinfo {title} {Direct observation of negative-index microwave surface waves},\ }\href {https://doi.org/10.1038/srep22018} {\bibfield  {journal} {\bibinfo  {journal} {Sci. Rep.}\ }\textbf {\bibinfo {volume} {6}},\ \bibinfo {pages} {22018} (\bibinfo {year} {2016})}\BibitemShut {NoStop}%
\bibitem [{\citenamefont {Yang}\ \emph {et~al.}(2017)\citenamefont {Yang}, \citenamefont {Jing}, \citenamefont {Shen}, \citenamefont {Wang}, \citenamefont {Zheng}, \citenamefont {Wang}, \citenamefont {Li}, \citenamefont {Shen}, \citenamefont {Koschny}, \citenamefont {Soukoulis},\ and\ \citenamefont {Chen}}]{yang2017hyperbolic}%
  \BibitemOpen
  \bibfield  {author} {\bibinfo {author} {\bibfnamefont {Y.}~\bibnamefont {Yang}}, \bibinfo {author} {\bibfnamefont {L.}~\bibnamefont {Jing}}, \bibinfo {author} {\bibfnamefont {L.}~\bibnamefont {Shen}}, \bibinfo {author} {\bibfnamefont {Z.}~\bibnamefont {Wang}}, \bibinfo {author} {\bibfnamefont {B.}~\bibnamefont {Zheng}}, \bibinfo {author} {\bibfnamefont {H.}~\bibnamefont {Wang}}, \bibinfo {author} {\bibfnamefont {E.}~\bibnamefont {Li}}, \bibinfo {author} {\bibfnamefont {N.-H.}\ \bibnamefont {Shen}}, \bibinfo {author} {\bibfnamefont {T.}~\bibnamefont {Koschny}}, \bibinfo {author} {\bibfnamefont {C.~M.}\ \bibnamefont {Soukoulis}},\ and\ \bibinfo {author} {\bibfnamefont {H.}~\bibnamefont {Chen}},\ }\bibfield  {title} {\bibinfo {title} {Hyperbolic spoof plasmonic metasurfaces},\ }\href {https://doi.org/10.1038/am.2017.158} {\bibfield  {journal} {\bibinfo  {journal} {NPG Asia Mater.}\ }\textbf {\bibinfo {volume} {9}},\ \bibinfo {pages} {e428} (\bibinfo {year} {2017})}\BibitemShut {NoStop}%
\bibitem [{\citenamefont {Yermakov}\ \emph {et~al.}(2018)\citenamefont {Yermakov}, \citenamefont {Hurshkainen}, \citenamefont {Dobrykh}, \citenamefont {Kapitanova}, \citenamefont {Iorsh}, \citenamefont {Glybovski},\ and\ \citenamefont {Bogdanov}}]{yermakov2018experimental}%
  \BibitemOpen
  \bibfield  {author} {\bibinfo {author} {\bibfnamefont {O.~Y.}\ \bibnamefont {Yermakov}}, \bibinfo {author} {\bibfnamefont {A.~A.}\ \bibnamefont {Hurshkainen}}, \bibinfo {author} {\bibfnamefont {D.~A.}\ \bibnamefont {Dobrykh}}, \bibinfo {author} {\bibfnamefont {P.~V.}\ \bibnamefont {Kapitanova}}, \bibinfo {author} {\bibfnamefont {I.~V.}\ \bibnamefont {Iorsh}}, \bibinfo {author} {\bibfnamefont {S.~B.}\ \bibnamefont {Glybovski}},\ and\ \bibinfo {author} {\bibfnamefont {A.~A.}\ \bibnamefont {Bogdanov}},\ }\bibfield  {title} {\bibinfo {title} {Experimental observation of hybrid {TE-TM} polarized surface waves supported by a hyperbolic metasurface},\ }\href {https://doi.org/10.1103/PhysRevB.98.195404} {\bibfield  {journal} {\bibinfo  {journal} {Phys. Rev. B}\ }\textbf {\bibinfo {volume} {98}},\ \bibinfo {pages} {195404} (\bibinfo {year} {2018})}\BibitemShut {NoStop}%
\bibitem [{\citenamefont {Won}(2019)}]{won2019into}%
  \BibitemOpen
  \bibfield  {author} {\bibinfo {author} {\bibfnamefont {R.}~\bibnamefont {Won}},\ }\bibfield  {title} {\bibinfo {title} {Into the {‘Mie-tronic’era}},\ }\href@noop {} {\bibfield  {journal} {\bibinfo  {journal} {Nat. Photon.}\ }\textbf {\bibinfo {volume} {13}},\ \bibinfo {pages} {585} (\bibinfo {year} {2019})}\BibitemShut {NoStop}%
\bibitem [{\citenamefont {Kivshar}(2022)}]{kivshar2022rise}%
  \BibitemOpen
  \bibfield  {author} {\bibinfo {author} {\bibfnamefont {Y.}~\bibnamefont {Kivshar}},\ }\href@noop {} {\bibinfo {title} {The rise of {Mie}-tronics}} (\bibinfo {year} {2022})\BibitemShut {NoStop}%
\bibitem [{\citenamefont {Koshelev}\ and\ \citenamefont {Kivshar}(2020)}]{koshelev2020dielectric}%
  \BibitemOpen
  \bibfield  {author} {\bibinfo {author} {\bibfnamefont {K.}~\bibnamefont {Koshelev}}\ and\ \bibinfo {author} {\bibfnamefont {Y.}~\bibnamefont {Kivshar}},\ }\bibfield  {title} {\bibinfo {title} {Dielectric resonant metaphotonics},\ }\href@noop {} {\bibfield  {journal} {\bibinfo  {journal} {ACS Photonics}\ }\textbf {\bibinfo {volume} {8}},\ \bibinfo {pages} {102} (\bibinfo {year} {2020})}\BibitemShut {NoStop}%
\bibitem [{\citenamefont {Koshelev}\ \emph {et~al.}(2020)\citenamefont {Koshelev}, \citenamefont {Kruk}, \citenamefont {Melik-Gaykazyan}, \citenamefont {Choi}, \citenamefont {Bogdanov}, \citenamefont {Park},\ and\ \citenamefont {Kivshar}}]{koshelev2020subwavelength}%
  \BibitemOpen
  \bibfield  {author} {\bibinfo {author} {\bibfnamefont {K.}~\bibnamefont {Koshelev}}, \bibinfo {author} {\bibfnamefont {S.}~\bibnamefont {Kruk}}, \bibinfo {author} {\bibfnamefont {E.}~\bibnamefont {Melik-Gaykazyan}}, \bibinfo {author} {\bibfnamefont {J.-H.}\ \bibnamefont {Choi}}, \bibinfo {author} {\bibfnamefont {A.}~\bibnamefont {Bogdanov}}, \bibinfo {author} {\bibfnamefont {H.-G.}\ \bibnamefont {Park}},\ and\ \bibinfo {author} {\bibfnamefont {Y.}~\bibnamefont {Kivshar}},\ }\bibfield  {title} {\bibinfo {title} {Subwavelength dielectric resonators for nonlinear nanophotonics},\ }\href@noop {} {\bibfield  {journal} {\bibinfo  {journal} {Science}\ }\textbf {\bibinfo {volume} {367}},\ \bibinfo {pages} {288} (\bibinfo {year} {2020})}\BibitemShut {NoStop}%
\end{thebibliography}%

\end{document}